\documentclass[10pt,twocolumn,english,prd,superscriptaddress,nofootinbib,preprintnumbers,showpacs,floatfix]{revtex4-2}

\usepackage{graphicx,color}
\usepackage{amsmath}
\usepackage{amssymb}
\usepackage{hyperref}
\usepackage[utf8]{inputenc}
\usepackage[english]{babel}
\usepackage{epsfig}
\usepackage{subfigure}
\usepackage{wasysym}
\usepackage{color,xcolor}
\usepackage{amsmath}
\usepackage{bm}
\usepackage{epsfig}
\usepackage{amsfonts}
\usepackage{dcolumn}
\usepackage{float}
\usepackage{booktabs}
\usepackage{relsize}

\hypersetup{colorlinks=true, linkcolor=blue, citecolor=green}

\usepackage{dcolumn}
\usepackage{bm}
\usepackage{ifpdf}
\usepackage{hyperref}
\usepackage{float}
\usepackage{bm}
\usepackage{xcolor,color,graphicx,graphics}
\usepackage[OT1]{fontenc}
\usepackage{latexsym,amssymb,amsmath,amsfonts}
\usepackage{makeidx}
\usepackage{epsfig}
\usepackage{epstopdf}
\usepackage{mathrsfs}
\hypersetup{colorlinks=true, linkcolor=blue, citecolor=green}
\usepackage{enumerate}
\usepackage{xcolor}
 \usepackage{multirow}
 \usepackage{tikz}

\newcommand{\orcidicon}{%
	\begin{tikzpicture}
		\draw[lime, fill=lime] (0,0) 
		circle [radius=0.16] 
		node[white] {{\fontfamily{qag}\selectfont \tiny ID}};
		\draw[white, fill=white] (-0.0625,0.095) 
		circle [radius=0.007];
	\end{tikzpicture}	\hspace{-2mm}
}
\newcommand\orcidEdnaldo{{\href{https://orcid.org/0000-0001-7230-3666}{\orcidicon}}}
\newcommand\orcidFrancisco{{\href{https://orcid.org/0000-0002-9388-8373}{\orcidicon}}}
\newcommand\orcidManuel{{\href{https://orcid.org/0000-0001-8586-0285}{\orcidicon}}}
\newcommand\orcidTarciso{{\href{https://orcid.org/0009-0007-0450-2672}{\orcidicon}}}
\newcommand\orcidHenrique{{\href{https://orcid.org/0000-0001-7565-4277}{\orcidicon}}}
\newcommand\orcidLuis{{\href{https://orcid.org/0009-0009-4322-6484}{\orcidicon}}}

\newcommand\orcidJorde{{\href{https://orcid.org/0009-0001-3344-2986}{\orcidicon}}}

\newcommand\orcidDiego{{\href{https://orcid.org/0000-0003-3984-9864}{\orcidicon}}}

\begin{document}

\title{Black bounce solutions in a realistic dark matter halo from M60*}

	\author{Ednaldo L. B. Junior\orcidEdnaldo\!\!} \email{ednaldobarrosjr@gmail.com}
\affiliation{Faculdade de F\'{i}sica, Universidade Federal do Pará, Campus Universitário de Tucuruí, CEP: 68464-000, Tucuruí, Pará, Brazil}
\affiliation{Programa de P\'{o}s-Gradua\c{c}\~{a}o em F\'{i}sica, Universidade Federal do Sul e Sudeste do Par\'{a}, 68500-000, Marab\'{a}, Par\'{a}, Brazil}

	\author{Jos\'{e} Tarciso S. S. Junior\orcidTarciso\!\!}
 \email{tarcisojunior17@gmail.com}
\affiliation{Faculdade de F\'{\i}sica, Programa de P\'{o}s-Gradua\c{c}\~{a}o em 
F\'isica, Universidade Federal do 
 Par\'{a},  66075-110, Bel\'{e}m, Par\'{a}, Brazil}

	\author{Francisco S. N. Lobo\orcidFrancisco\!\!} \email{fslobo@ciencias.ulisboa.pt}
\affiliation{Instituto de Astrof\'{i}sica e Ci\^{e}ncias do Espa\c{c}o, Faculdade de Ci\^{e}ncias da Universidade de Lisboa, Edifício C8, Campo Grande, P-1749-016 Lisbon, Portugal}
\affiliation{Departamento de F\'{i}sica, Faculdade de Ci\^{e}ncias da Universidade de Lisboa, Edif\'{i}cio C8, Campo Grande, P-1749-016 Lisbon, Portugal}

	\author{Jorde A. A. Ramos\orcidJorde\!\!}
 \email{jordefisica@gmail.com}
\affiliation{Faculdade de F\'{\i}sica, Programa de P\'{o}s-Gradua\c{c}\~{a}o em 
F\'isica, Universidade Federal do 
 Par\'{a},  66075-110, Bel\'{e}m, Par\'{a}, Brazil}
 
	\author{Manuel E. Rodrigues\orcidManuel\!\!}
	\email{esialg@gmail.com}
	\affiliation{Faculdade de F\'{\i}sica, Programa de P\'{o}s-Gradua\c{c}\~{a}o em 
F\'isica, Universidade Federal do 
 Par\'{a},  66075-110, Bel\'{e}m, Par\'{a}, Brazil}
\affiliation{Faculdade de Ci\^{e}ncias Exatas e Tecnologia, 
Universidade Federal do Par\'{a}\\
Campus Universit\'{a}rio de Abaetetuba, 68440-000, Abaetetuba, Par\'{a}, 
Brazil}

\author{Diego Rubiera-Garcia\orcidDiego\!\!} \email{ drubiera@ucm.es}
\affiliation{Departamento de Física Téorica and IPARCOS, Universidad Complutense de Madrid, E-28040 Madrid, Spain}

\author{Luís F. Dias da Silva\orcidLuis\!\!} 
        \email{fc53497@alunos.fc.ul.pt}
\affiliation{Instituto de Astrof\'{i}sica e Ci\^{e}ncias do Espa\c{c}o, Faculdade de Ci\^{e}ncias da Universidade de Lisboa, Edifício C8, Campo Grande, P-1749-016 Lisbon, Portugal}


 \author{Henrique A. Vieira\orcidHenrique\!\!} \email{henriquefisica2017@gmail.com}
\affiliation{Faculdade de F\'{i}sica, Programa de P\'{o}s-Gradua\c{c}\~{a}o em F\'{i}sica, Universidade Federal do Par\'{a}, 66075-110, Bel\'{e}m, Par\'{a}, Brazil}


\begin{abstract}
We formulate a Simpson--Visser black bounce solution embedded in a dark matter halo. The latter is modeled using an empirical density profile calibrated from observations of the elliptical galaxy NGC 4649 (M60), based on imaging from the Hubble Space Telescope, stellar velocity dispersion data, and the dynamics of globular clusters. The resulting space-time metric, in addition to retaining dependence on the mass parameter $m$, the asymptotic circular velocity $V_c$, and the halo scale radius $a$, also depends on the regularization parameter $q_H$. It reduces to the canonical black bounce solution without a halo in the limit $V_c\to0$ (or $a\to\infty$), and to the Schwarzschild solution with a dark matter halo when $q_H\to0$. We analyze the response of fundamental geometrical and physical quantities in the presence of the halo, such as the event horizon radius, the shadow size, and some curvature invariants. In particular, we show that the observational range of the shadow radius, as inferred from the imaging of Sagittarius A*, constrains the parameter space of the solution to regular black hole configurations, excluding wormhole scenarios. We also study the dynamics of massless particles in this scenario through the effective potential and examine thermodynamic properties, highlighting the impact on thermodynamic potentials in terms of entropy. Finally, we extend the analysis to scenarios involving electromagnetic fields non-minimally coupled to a phantom scalar field, considering configurations with either purely magnetic or purely electric charge. Our results suggest that the dark matter halo influences both the internal geometry and the observational properties of black bounces, imposing constraints on the solution's parameter space based on astrophysical data. This highlights the importance of incorporating astrophysical environments into the modeling of regular black holes and wormholes, offering new avenues for testing gravity in the strong-field regime.
\end{abstract}
\date{\today}
\maketitle
\def\HMS{{\scriptscriptstyle{\rm HMS}}}

\section{Introduction}\label{sec1}

Formulated by Albert Einstein in the early twentieth century, the theory of General Relativity (GR) \cite{Einstein:1916vd} is the currently established description of gravity. Karl Schwarzschild \cite{Schwarzschild:1916uq} found the first (vacuum) solution of its field equations, describing an object with a casual boundary, the event horizon, and which is characterized by mass alone, currently understood as a black hole (BH). The inclusion of spin provides the astrophysically relevant solution, namely, the Kerr black hole \cite{Kerr:1963ud}, and when charge is added then we find the Kerr-Newman BH \cite{Newman1965,Null-temp012}. As a result of technological advances in high-precision astrophysical measurements, interest in BHs has greatly increased in the last decades, leading to milestones such as the direct detection of gravitational waves by the LIGO Scientific Collaboration and Virgo Collaboration \cite{LIGOScientific:2016aoc,LIGOScientific:2017ync}, originating from BHs mergers, as well as by the event-horizon imaging of supermassive black holes (SBHs) at the centers of M87 and the Milky Way galaxies, as obtained by the Event Horizon Telescope \cite{EventHorizonTelescope:2019dse,EventHorizonTelescope:2022wkp}.

Despite their astrophysical plausibility, there are significant challenges towards their complete understanding. One of the most notable ones is the unavoidable presence of a space-time singularity deep in their innards. It is defined by the presence of incomplete geodesics curves at some region is approached ($r=0$ in Schwarzschild case, and the ring singularity in Kerr) \cite{Senovilla:2014gza}, and is typically correlated with the divergence of some sets of curvature scalars. Attempts to solve this problem have ranged far and wide, in particular, via the incorporation of additional fields coupled to the Einstein field equations, such as electromagnetic fields. This is motivated by the possibility of finding regular black holes (RBHs) which are free of curvature singularities \cite{Ansoldi:2008jw}. This was pioneered by James Bardeen, whose model was introduced in 1968 \cite{Bardeen}, later interpreted as a solution of Einstein field equations coupled to nonlinear electrodynamics (NLED) and a magnetic monopole source \cite{Ayon-Beato:2000mjt}, but also in terms of an electric charge  \cite{Rodrigues:2018bdc}.
A wide range of other RBH solutions supported by NEDs has been developed in the literature, see e.g. \cite{Bronnikov:2000vy,Dymnikova:2004zc,Balart:2014cga,Culetu:2014lca}. The presence of NLED influences BH properties, such as particle dynamics \cite{Novello:1999pg,Habibina:2020msd,Toshmatov:2021fgm,dePaula:2024yzy}, the features of BH images \cite{Stuchlik:2019uvf,Allahyari:2019jqz,Kruglov:2020tes}, and their thermodynamic properties \cite{Breton:2004qa,Myung:2007xd,Ma:2015gpa,Kruglov:2016ymq,Fan:2016hvf,Barbagallo:2026hof}.

Within this context, a proposal by Alex Simpson and Matt Visser in 2018 \cite{Simpson:2018tsi} introduced singularity-free solutions known as black bounces (BB). The core idea is to avoid the central singularity via the promotion of the radial coordinate to a radial function following the prescription of Ellis' wormhole \cite{Ellis:1973yv}
$r\to\sqrt{r^2+q^2}$, where $q$ is a new parameter. This generalization encompassing two possible scenarios depending on the value of $q$: the first corresponds to RBHs while the second describes traversable wormholes (WHs), both of one-way and two-way types. WHs describe space-times featuring a throat instead of the central singularity and they have been subject of considerable interest for decades \cite{Bronnikov:1973fh,Morris:1988cz,Barcelo:1999hq,Barcelo:2000zf,Visser:2003yf,Lobo:2005us,Lobo:2007zb,Cardoso:2016rao,Bronnikov:2017sgg,Lobo:2017cay,Blazquez-Salcedo:2020czn,Churilova:2021tgn,Konoplya:2021hsm}. As for BB space-times, they have motivated the investigation of their causal structures and energy conditions \cite{Lobo:2020ffi} or gravitational lensing effects \cite{Nascimento:2020ime,Tsukamoto:2020bjm,Cheng:2021hoc,Tsukamoto:2021caq,Zhang:2022nnj}, and have been generalized to rotating models \cite{Mazza:2021rgq,Xu:2021lff}. Furthermore, a general formalism for the systematic construction of BB-type solutions, assuming spherical symmetry and staticity, was reported in \cite{Alencar:2025jvl}.

The theoretical support of BB-type solutions is commonly implemented through the inclusion of NLED and scalar fields as matter sources \cite{Bronnikov:2021uta}. However, such an interpretation introduces some drawbacks, such as the immediate suppression of NLED effects outside the vicinity of the compact object \cite{Mignani:2016fwz,Ejlli:2020yhk}, thereby failing to provide an appropriate transition to asymptotically flat regions where weak electromagnetic fields, described by Maxwell’s theory, prevail. This may have significant implications for astrophysical observations, suggesting the need to consider the effective metric in photon propagation \cite{Novello:1999pg}, where such metrics may fail to preserve fundamental properties, such as regularity of propagation. 

Within this context, recently some of us presented a formulation of BB-type solutions that considers linear electrodynamic (LED) scenarios with an interaction term of the form $W(\varphi)\mathcal{L}(\mathcal{F})$ \cite{bounceLED}. This allows for a systematic reconstruction of the matter functions that characterize the space-time geometry, resulting in global regularity of the solution. Although the solutions reconstructed this way address various aspects of isolated compact objects, it does not account for interactions with an external matter component. One particularly relevant scenario of such matter environments around such compact objects is the one supplied by dark matter (DM). 

The existence of DM has been inferred main by galactic rotation curves \cite{bertone2005,KFreese2008,Swart2017,Risa2018,Arbey2021} (see \cite{cebrian2022,Misiaszek2023} for alternative methods), and it is currently estimated to account for $\sim 5/6$ of the mass of the Universe \cite{Planck2021}. Alongside other forms of matter, DM may act in the formation of galaxies within a DM halo \cite{Valluri2003}. In fact, there are galaxies that host BHs at their centers \cite{M87SBH2019,SagittariusA2023}, motivating studies on the possible interactions between these objects and the DM halos in which they are immersed \cite{ElZant2020}. A typical DM halo exhibits a higher density at the center, becoming progressively more diffuse towards the outer regions \cite{TBSW1997-perfisDM}. Among DM profiles that aim to describe its distribution, there are some well used ones, such as the Hernquist profile \cite{Hernquist1990,Liu2022,Shen2024,Fard2024}, the Navarro-Frenk-White (NFW)  \cite{NFW1997,Tsuchiya2013,AA01}, or the Dehnen one \cite{DEHNEN1993, Otebay2021}. The modeling of DM halos via such profiles provides modifications to the orbits of the $S$-stars \cite{Xamidov,Heibel21},  accretion disk emission \cite{Nieto,Giambo}, or weak gravitational lensing \cite{Sucu49}. The influence of the halo may be small for typical parameters, ensuring a gradual transition to the case without DM \cite{Lobo2025H1,Lobo2025H2}. However, regimes with stronger parameters may produce measurable effects, such as alterations in the BH shadow, still compatible with observations of Sgr A* and M87* \cite{Eid}. The potential of such matter environments around compact objects to modify the observable signatures \cite{Fonseca:2025ehf}, for instance using extreme-mass ratio inspiral systems via future detectors such as the Square Kilometre Array (SKA) \cite{Guo26,Nicolini26}, or in $S$-stars via current (such as Keck and GRAVITY) of future observatories (such as the Thirty Meter Telescope (TMT) \cite{Kalita26}) motivates the present work.

The main aim of this work is to investigate BB-type solutions when embedded in a DM halo. To this end we shall employed the recently introduced DM density profile in \cite{mod1}, which provides a more complete description of the mass distribution in massive elliptical galaxies, illustrated with the case of NGC 4649 (M60). Using the corresponding density profile $\rho_{DM}$ and the parameter ranges of the halo presented there, we investigate the response of certain geometrical aspects of the space-time due to the presence of the halo, such as the BH shadow -- using observations from the EHT of Sgr A* \cite{shadow} --. Furthermore, we shall extend such an analysis to the context of  the framework introduced in \cite{bounceLED}, by which a purely magnetic or electric charge is implemented within a non-minimal coupling to a phantom scalar field. In all scenarios considered in this work, we shall deal with spherically symmetric and static solutions with metric signature $(+,-,-,-)$, as well as geometrized units $(G=c=1)$.

This work is organized as follows. In Section~\ref{sec2} we present the density profile, the datasets of the halo parameters employed, and the generalized metric. In Section~\ref{sec3} we analyze some initial geometrical properties, such as the event horizon, the Kretschmann scalar, and the shadow radius, followed by a brief treatment of the dynamics of a massless particle, and then some thermodynamic properties. In Section~\ref{sec:electrodynamic} we present the field equations of GR coupled to a scalar field and to an interaction term between the scalar field and LED, where we include the contribution of the DM halo. We briefly treat separately the magnetic and electric cases as sources of our BB-type solutions. Finally, conclusions are given in Section~\ref{sec:concl}.

\section{BLACK HOLE SOLUTION EMBEDDED IN A DARK MATTER HALO}\label{sec2}
\subsection{Dark matter density profile}
The DM density chosen in this work has the following radial dependence \cite{mod1}
\begin{equation}
\rho_{DM}(r)=\frac{V_c^2}{4\pi}\frac{\left(3a^2+r^2\right)}{\left(a^2+r^2\right)^2} \ ,
\end{equation}
using a Schwarzschild BH as the background geometry. In this model the density responds directly to the halo velocity and its radius, such that if the velocity vanishes, $V_c\to0$, of the halo is very diffuse, $a\to\infty$, the density vanishes. An extension of the Schwarzschild-type solution is addressed in \cite{Lobo2025H1}, indicating that the halo contributes directly to the metric function. Such a contribution can be removed in the limits $V_c\to0$ or $a\to\infty$.

We adopt the values defined in \cite{mod1}, associated with the parameters of mass $m$, critical velocity $V_c$, and critical radius $a$. These parameters define the ranges listed in Table~\ref{tab:parametros}, including both SI units and geometrized units \cite{Lobo2025H1}, for consistency and completeness. We consider two main datasets. The first dataset, which we refer to as {\bf Data I}, is based on X-ray observational data. The second dataset, referred to as {\bf Data II}, arises from the modelling of the dynamical mass profile. For each dataset, the following values are presented:

\begin{itemize}
	\item \textbf{Data I}: $V_{c}^{(I)} = 13.68 \times 10^{-4}$, $a^{(I)} = 30.86 \times 10^{19}\,\text{m}$, and $M^{(I)} = 5.17 \times 10^{12}\,\text{m}$;
	\item \textbf{Data II}: $V_{c}^{(II)} = 18.35 \times 10^{-4}$, $a^{(II)} = 46.29 \times 10^{19}\,\text{m}$, and $M^{(II)} = 6.65 \times 10^{12}\,\text{m}$.
\end{itemize}
These datasets serve as input values for our numerical and analytical analyses of RBHs and WHs immersed in DM haloes. Additionally, as shown in \cite{mod1}, the mass of the BH in the absence of a DM halo is given by $M_0 = 4.3 \times 10^{9} M_{\odot}$, which corresponds to $M_0 = 6.35 \times 10^{12}\,$m in geometrized units. Another additional parameter relevant for investigating the properties of the solution is the distance between the observer (on Earth) and the galaxy M60, given by $r_O = 15.7 \times 10^{3}\,$kpc, which is equivalent to $r_O = 48.45 \times 10^{22}\,$m.

\begin{table}[t!]
    \centering
\renewcommand{\arraystretch}{1.2}
    \begin{tabular}{lcc}
        \toprule
        \textbf{Parameter} & \textbf{Minimum} & \textbf{Maximum} \\
        \midrule
        $V_c$[km/s] (IS) & $300$ & $1100$ \\
        $V_c$ (Geom.) & $10 \times 10^{-4}$ & $36.7 \times 10^{-4}$ \\
        $a$[kpc] (IS) & $5$ & $90$ \\
        $a$[m] (Geom.) & $15.43 \times 10^{19}$ & $277.74 \times 10^{19}$ \\
        $M$[$M_{\odot}$] (IS) & $2.5 \times 10^{9}$ & $6 \times 10^{9} $ \\
        $M$[m] (Geom.) & $3.69 \times 10^{12}$ & $8.86 \times 10^{12}$ \\
        \bottomrule
    \end{tabular}
    \caption{Parameters $V_c$, $a$ and $M$ in different units with minimum and maximum values. The values in SI were taken from \cite{mod1}.}
    \label{tab:parametros}
\end{table}

\subsection{Spacetime metric}
For the solutions obtained in this work, we assume a spherically symmetric and static solution presented in the following general metric form:
\begin{equation}
ds^2=A(r)dt^2-\frac{1}{A(r)}dr^2-r^2d\Omega^2 \ ,\label{metrica_geral}
\end{equation}
where the angular surface element in spherical coordinates is given by $d\Omega^2 = d\theta^2 + \sin^2{\theta}\,d\phi^2$. We start from the well known Schwarzschild-type BH solution described by the metric coefficient
\begin{equation}
    A_{\rm schw}(r)=1-\frac{2M_0}{r}\ ,\label{sol_schw}
\end{equation}
where the radius of the event horizon is $R=2M_0$, where $M_0$ corresponds to the mass of the solution without a halo. 

As discussed in the introduction, in the transition from the Schwarzschild  solution to the BB one follows the replacement $r\to\sqrt{r^2+q^2}$, such that in the limit $q\to0$ the Schwarzschild solution is recovered. In this way, the BB metric reads as
\begin{equation}
    ds^2=A(r)dt^2-\frac{1}{A(r)}dr^2-\Sigma(r)^2d\Omega^2 \ ,\label{metrica_geral_BB}
\end{equation}
where $m$ is the mass parameter, the radial function reads as $\Sigma(r)=\sqrt{r^2+q^2}$ with $r \in(-\infty,+\infty)$ a radial parameter, and
\begin{equation}
    A_{\rm BB}(r)=1-\frac{2m}{\sqrt{r^2+q^2}}\ ,\label{sol_BB}
\end{equation}
Here $q$ is the parameter that determines the nature of the solution: a regular black hole (RBH) if $0<q<2m$, a one-way wormhole (one-WH) if $q=2m$, and a two-way wormhole (two-WH) if $q>2m$. (a) For the first case, the RBH exhibits two horizons $r_h\pm = \pm \sqrt{4m^2 - q^2}$, with a bounce located at $r = 0$ ($\Sigma(r=0)=q$), understood as boundary between two space-time patches, described as a space-like spherical surface. (b) The second case corresponds to a WH with a null extremal throat located at $r = 0$, which is traversable only in one direction. (c) The third case describes a WH with a bidirectional throat, being traversable in both directions, with a time-like throat at $r = 0$.

Following this phenomenology, we consider the following metric function corresponding to the Schwarzschild solution with a DM halo \cite{Lobo2025H1}, given by
\begin{equation}
    A_{\rm schw,H}^{\rm}(r)=1-\frac{2m}{r}-\frac{2V_c^2r^2}{(a^2+r^2)}\label{sol_halo_schw}\ ,
\end{equation}
which reads, after the implementation of the BB replacement, $r\to\Sigma(r) =\sqrt{r^2 + q_{H}^2}$, as
\begin{equation}
    A(r)=1-\frac{2m}{\sqrt{r^2+q_{H}^2}}-\frac{2V_c^2(r^2+q_{H}^2)}{(a^2+r^2+q_{H}^2)}\label{sol_halo_BB}\ .
\end{equation}
This corresponds to a BB-type BH solution with a DM halo, where $q_H$ denotes the BB parameter for a solution with a halo. Obviously, when $V_c = 0$, then the usual BB-type solution of Eq.~(\ref{sol_BB}) is recovered.

\section{Geometric and Thermodynamic Properties}\label{sec3}

\subsection{Event Horizon}\label{subsec:event_horizon}

From Eq.~(\ref{sol_halo_BB}), we set $A(r) = 0$ in order to obtain the event horizons of the solution (on each side of the throat), for the values of Data I and II
\begin{equation}
r_{hI\pm}= \pm 0.739\sqrt{1.9268\times10^{26}-1.831q_H^2}\label{rhI} \ ,
\end{equation}
and
\begin{equation}
r_{hII\pm}= \pm 0.2887\sqrt{2.0552\times10^{27}-11.9998q_H^2} \ , \label{rhII}
\end{equation}
respectively. The existence of event horizons (either under the form of two distinct horizons or as single degenerate horizon at the throat), imposes constraints on the parameter $q_H$, with $[q_H]=m$. From Eqs.~(\ref{rhI})~and~(\ref{rhII}), the following intervals are obtained
\begin{equation}
-1.02581\times10^{13}\leq q_H\leq 1.02581\times10^{13}\ ,
\end{equation}
from $r_{hI}$, and 
\begin{equation}
-1.30871\times10^{13}\leq q_H\leq 1.30871\times10^{13}\ ,
\end{equation}
from $r_{hII}$. Returning to the BB solution without a halo, from Eq.~(\ref{sol_BB}), we have that $q$ can likewise be constrained by the event horizons $r_{h\pm}$, as
\begin{equation}
-1.27\times10^{13}\leq q \leq1.27\times10^{13} \ .
\end{equation}
This indicates that the presence of the halo implies a modification of the domain of $q \to q_H$, with its range increasing when the BH mass grows, $M_0 \to M_{II}$, or decreasing when the mass diminishes, $M_0 \to M_I$.

\begin{figure}[t!]
\includegraphics[width=\linewidth]
{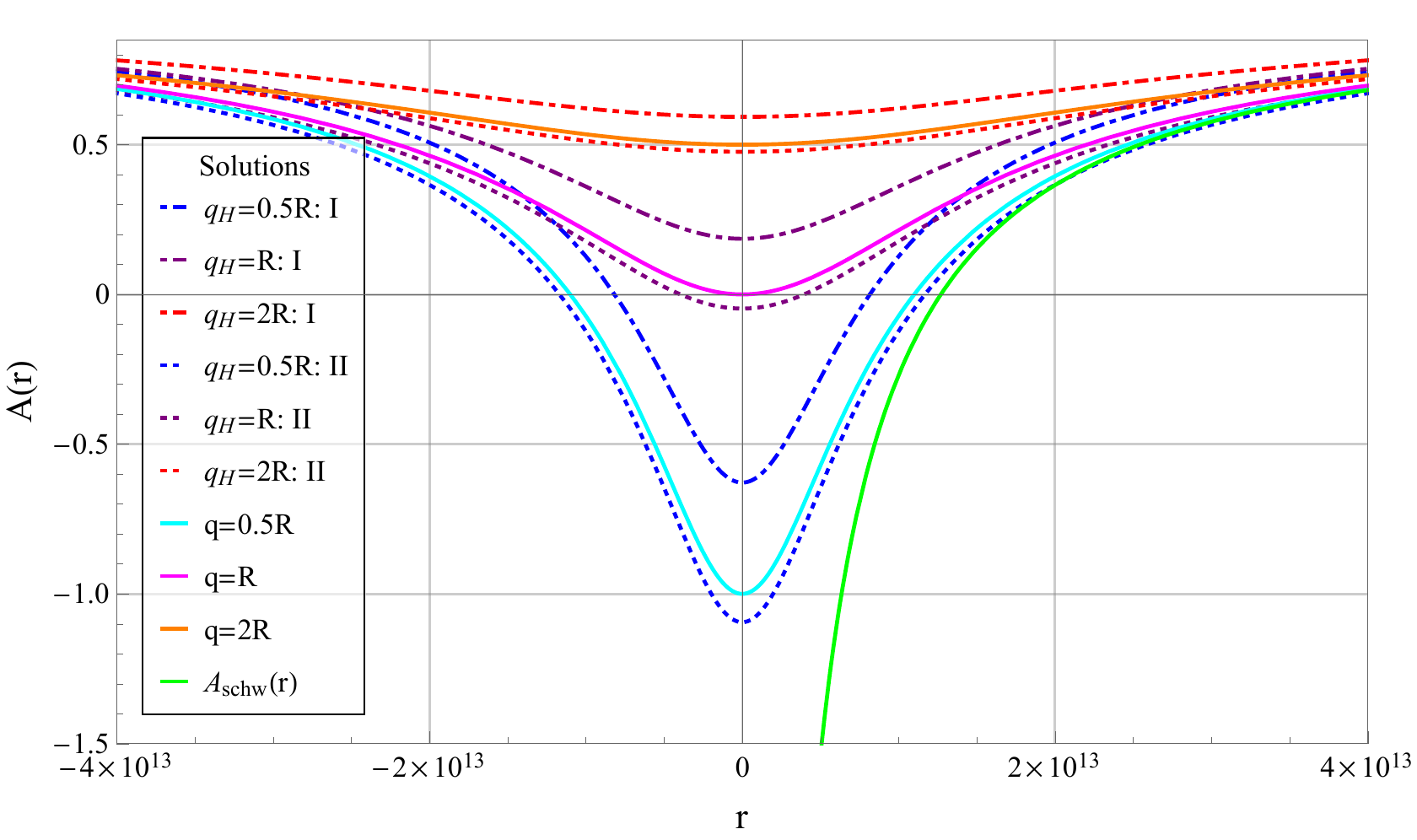}
\caption{Graphical representation of the metric function $A(r)$ for the following solutions: BB with a halo from Eq.~(\ref{sol_halo_BB}), BB without a halo from Eq.~(\ref{sol_BB}), and Schwarzschild without a halo from Eq.~(\ref{sol_schw}). We extend the BB solution with a halo to the datasets of the parameters $(V_c, a)$ from Table~\ref{tab:parametros}, denoted as I, for $(M_I, V_{cI}, a_I)$, and II, for $(M_{II}, V_{cII}, a_{II})$. The values $q_H = (0.5R, R, 2R)$ were adopted in terms of the Schwarzschild event horizon radius $R = 2M_0$.
} 
\label{fig:F(r)_vs_r_model02}
\end{figure}

Fig.~\ref{fig:F(r)_vs_r_model02} shows the function $A(r)$ from Eq.~(\ref{sol_halo_BB}), illustrating that the existence and number of event horizons preserve the solution’s dependence on the values assigned to $q_H$, now taking into account the inclusion of the halo in the solution. We consider three main scenarios, with $q = q_H = (0.5R, R, 2R)$, where $R = 2M_0$. For these three cases the following behaviour is observed:

\begin{itemize}

\item For $q = q_H = 0.5R$, even with the halo contribution, two event horizons $r_{hI,II\pm}$ are still defined. However, their values vary in response to changes in the BH mass due to the parameters $(V_c, a)$ characterizing the presence of the halo in the solution. Specifically, for Data I there is a reduction in the pair of horizons $r_{hI\pm}$, corresponding to the decrease in BH mass $(M_I < M_0)$ when the halo is included. For Data II, there is a slight increase in the pair of horizons $r_{hII\pm}$, reflecting the increase in BH mass due to the halo $(M_{II} > M_0)$. 

\item For $q = q_H = R$, upon inclusion of the halo, changes occur between the categories within the BB solution. Specifically, when assigning the values of Data I to the solution, the curve of the function shifts upwards, resulting in the absence of a horizon. In this regime, the presence of the halo affects the type of hypersurface at $r = 0$ that characterises the solution, causing it to transition from a one-WH to a two-WH, as what was previously a null throat becomes a bidirectional throat.

For the solution to be characterized solely as a one-WH, such that $A(r=0)=0$, it is necessary that
\begin{equation}
q_H=\pm1.034\times10^{13}[m]\equiv \pm R_{HI}   \ ,
\end{equation}
where $R_{HI}$ is the radius of the event horizon for the Schwarzschild solution with a halo from Eq.~(\ref{sol_halo_schw}) for Data I \cite{Lobo2025H1}. When assigning the values of Data II to this one-WH solution, the curve of the function $A(r)$ shifts downwards, resulting in the formation of two horizons. That is, in this regime of $q = q_H$, the BB solution without a halo, from Eq.~(\ref{sol_BB}), is characterized as a one-WH, but when the halo is included, from Eq.~(\ref{sol_halo_BB}), the solution becomes characterized as an RBH. In general, a transition one-WH $\to$ RBH occurs.

On the other hand, using Data II, for the solution to correspond to a one-WH it is necessary that
\begin{equation}
     q_H=\pm1.33\times10^{13}[m]\equiv \pm R_{HII}   \ ,
\end{equation}
where $R_{HII}$ is the radius of the event horizon for the Schwarzschild solution with a halo from Eq.~(\ref{sol_halo_schw}).

\item For $q = q_H = 2R$, there is a shift in the curve of the function at the point $A(r=0)$ due to the contribution of the halo, such that
\begin{equation}
    1-\frac{2M_0}{q}\to1-\frac{2M_{I,II}}{q_H}-\frac{2q_H^2V_c^2}{a^2+q_H^2} \ ,
\end{equation}
being shifted upwards for Data I, and downwards for Data II. 

\end{itemize}

In order for the coordinate $t$ to remain timelike, such that the hypersurface located at $r = 0$ remains timelike ($g_{00} > 0 \to A(r=0) > 0$), it is necessary that
\begin{equation}
\left(1-\frac{2M_{I,II}}{q_H}-\frac{2q_H^2V_c^2}{a^2+q_H^2}\right)>0 \ ,
\end{equation}
implying that
\begin{equation}
-1.034\times10^{13}[m]\ < q_H < 1.034\times10^{13}[m] \ ,  
\end{equation} 
for Data I, where
 $1.034\times10^{13}[m]= R_{HI}$, and
\begin{eqnarray}
    -1.33\times10^{13}[m] < q_H < 1.33\times10^{13}[m] \ ,
\end{eqnarray}
for Data II, where $1.33\times10^{13}[m]= R_{HII}$.

In general, for both datasets, when comparing the BB solution without a halo, Eq.~(\ref{sol_BB}), with the BB solution with a halo, Eq.~(\ref{sol_halo_BB}), for the same value of $q = q_H$, we have that for the regime: $q_H < R$, there is a variation in the radius of the event horizons that follows the change in mass due to the presence of the halo; $q_H = R$, corresponding to the null throat of the WH, the halo causes a modification of the object, making it transition to a different type within the BB solution; $q_H > R$ results in a restriction of the domain of $q_H$ using the horizon of the Schwarzschild solution with a halo, $q_H > R_H$. This analysis suggests that, just as the horizon of the Schwarzschild solution without a halo, $R$, is the limiting value between classes in the BB solution without a halo, the horizon of the Schwarzschild solution with a halo, $R_H$, remains as the threshold between classes within the BB solution with a halo.

The regularizing parameter $q_H$ has its critical value associated with the existence of the event horizon, specifically when $q_H|_{r_h\to0}\to q_{H}^{crit}$. Its value is slightly modified in response to the halo, showing a difference of $q^{crit}/M_0-q_H^{crit}/M_{I,II}=1.776\times10^{-15}$, highlighting the sensitivity of the horizon structure to the contribution of the halo.

\subsection{Kretschmann Scalar}

To analyze the regularity of the above space-times with respect to the halo parameters, we shall use the Kretschmann scalar $K = R_{\mu\nu\alpha\beta}R^{\mu\nu\alpha\beta}$, where $R_{\mu\nu\alpha\beta}$ denotes the Riemann tensor. For the general metric of Eq.~(\ref{metrica_geral}), the scalar $K$ takes the form
\begin{equation}
    K = A''(r)^2+\frac{4 A'(r)^2}{r^2}+\frac{4
   [A(r)-1]^2}{r^4} \ ,
\end{equation}
which may be used for a Schwarzschild-type solution with or without a halo. For the metric in Eq.~(\ref{metrica_geral_BB}), where we use the solution from Eq.~(\ref{sol_halo_BB}), the scalar $K$ takes the form
\begin{eqnarray}
K &=& \frac{4}{\Sigma^4}\Big\{1+(4\Sigma^2+\Sigma'^4)(\eta_{BB}^2+\eta_H^2)+2r\Sigma^2(\eta'_{BB}
	\nonumber \\
	&&
-\eta'_H)^2+2(\eta_H-\eta_{BB})[2r+\Sigma^2\Sigma'\Sigma''(\eta'_H-\eta'_{BB})]
\nonumber \\
	&&
-2\eta_{BB}\eta_H(\Sigma'^4+4\Sigma^2)+\frac{\Sigma^4}{4}(\eta''_{BB}-\eta''_{H})^2\Big\}\ , \label{K_scalar_BB_halo}
\end{eqnarray}
where, for simplicity, we set $F(r)=\eta_{BB}(r)-\eta_{H}(r)$, $\eta_{BB}(r)=1-2m/\Sigma$ and $\eta_{H}(r)=2V_c^2\Sigma^2/(a^2+\Sigma^2)$, in which $\Sigma=\Sigma(r)$. The overlines indicate derivatives with respect to the radial coordinate.

By setting $V_c = a = 0$, we recover the Kretschmann scalar of the BB solution without a halo, $K_{BB}$, and if $V_c = a = q_H = 0$, we recover the Kretschmann scalar of the Schwarzschild solution without a halo, $K_{\rm schw}$; that is, $K(r)|_{(V_c=0,a=0)}=K_{BB}(r)$ and $K(r)|_{(V_c=0,a=0,q_H=0)}=K_{\rm schw}(r)$.
Asymptotically, the curvature follows the behaviour for Eq.~\eqref{K_scalar_BB_halo} given by
\begin{equation}
	\lim_{r \to 0} K = \frac{20m^2}{q_H^6} + \frac{16m V_c^2}{q_H^3(a^2+q_H^2)} + \frac{32V_c^4}{(a^2+q_H^2)^2}\ ,
	\label{limit_K_r=0}
\end{equation}
and
\begin{equation}
	\lim_{r \to \infty} K \rightarrow 0\ ,
\end{equation}
which characterizes flat spacetime as $r\to\infty$ and the regularity at the center of the object.

\begin{figure}[t!]
    \centering
    \begin{subfigure}
        \centering
        \includegraphics[width=\linewidth]{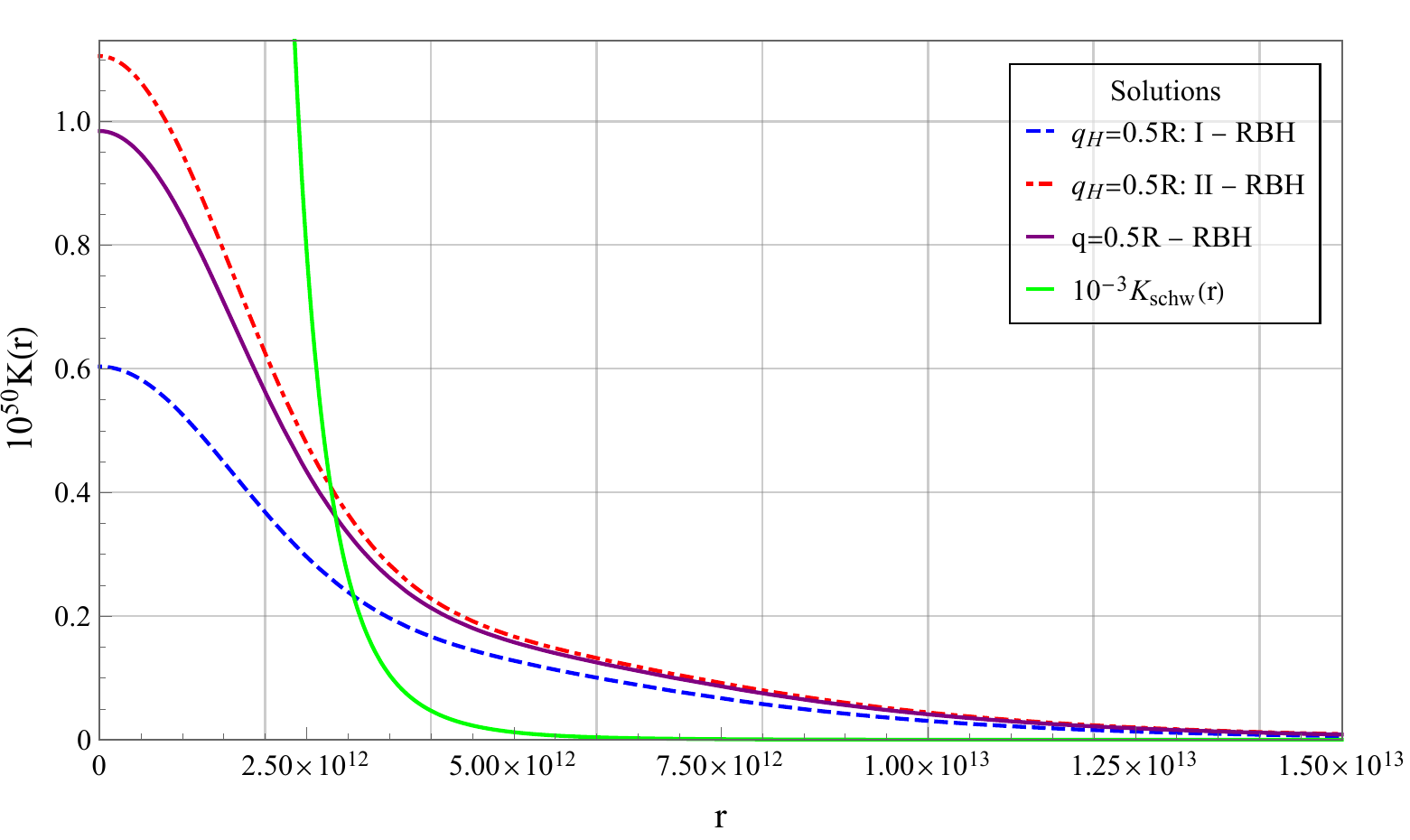}
    \end{subfigure}

    \begin{subfigure}
        \centering
        \includegraphics[width=\linewidth]{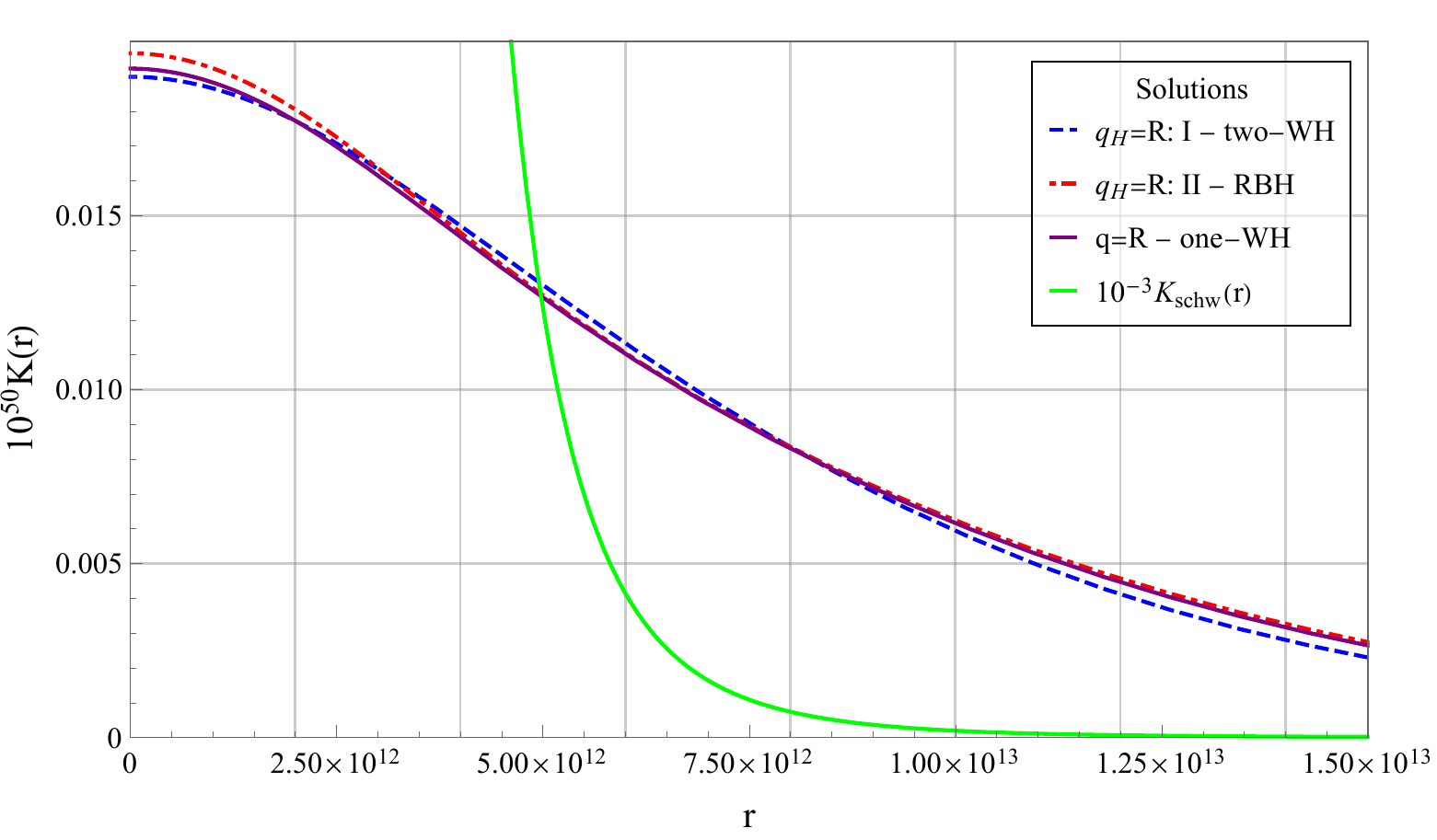}
    \end{subfigure}

    \begin{subfigure}
        \centering
        \includegraphics[width=\linewidth]{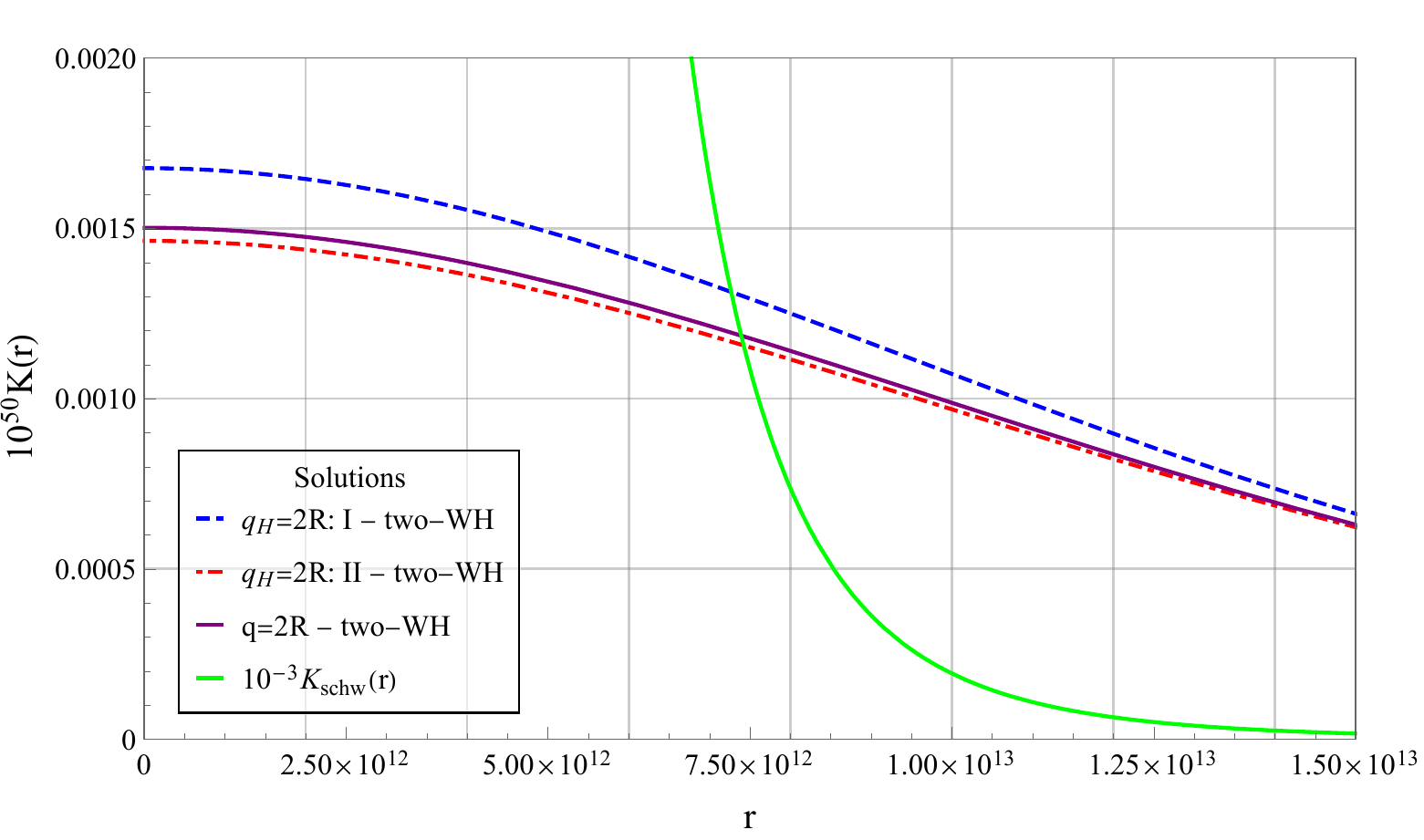}
    \end{subfigure}

    \caption{ Graphical representation of the evolution of the Kretschmann scalar $K(r)$ for the following solutions: BB with a halo from Eq.~(\ref{K_scalar_BB_halo}), BB without a halo ($K_{BB}$) from Eq.~(\ref{K_scalar_BB_halo}) with $V_c = a = 0$, and Schwarzschild without a halo ($K_{\rm schw}$) from Eq.~(\ref{K_scalar_BB_halo}) with $V_c = a = q_H = 0$. This is considered for different values of $q_H = (0.5R, R, 2R)$, where $R = 2M_0$ is the event horizon radius of the Schwarzschild solution. We extend the BB solution with a halo to the datasets of the parameters $(m, V_c, a)$ from Table~\ref{tab:parametros}, denoted as I, for $(M_I, V_{cI}, a_I)$, and II, for $(M_{II}, V_{cII}, a_{II})$.}
    \label{fig:scalarK}
\end{figure}

\begin{figure}[t!]
\includegraphics[width=\linewidth]
{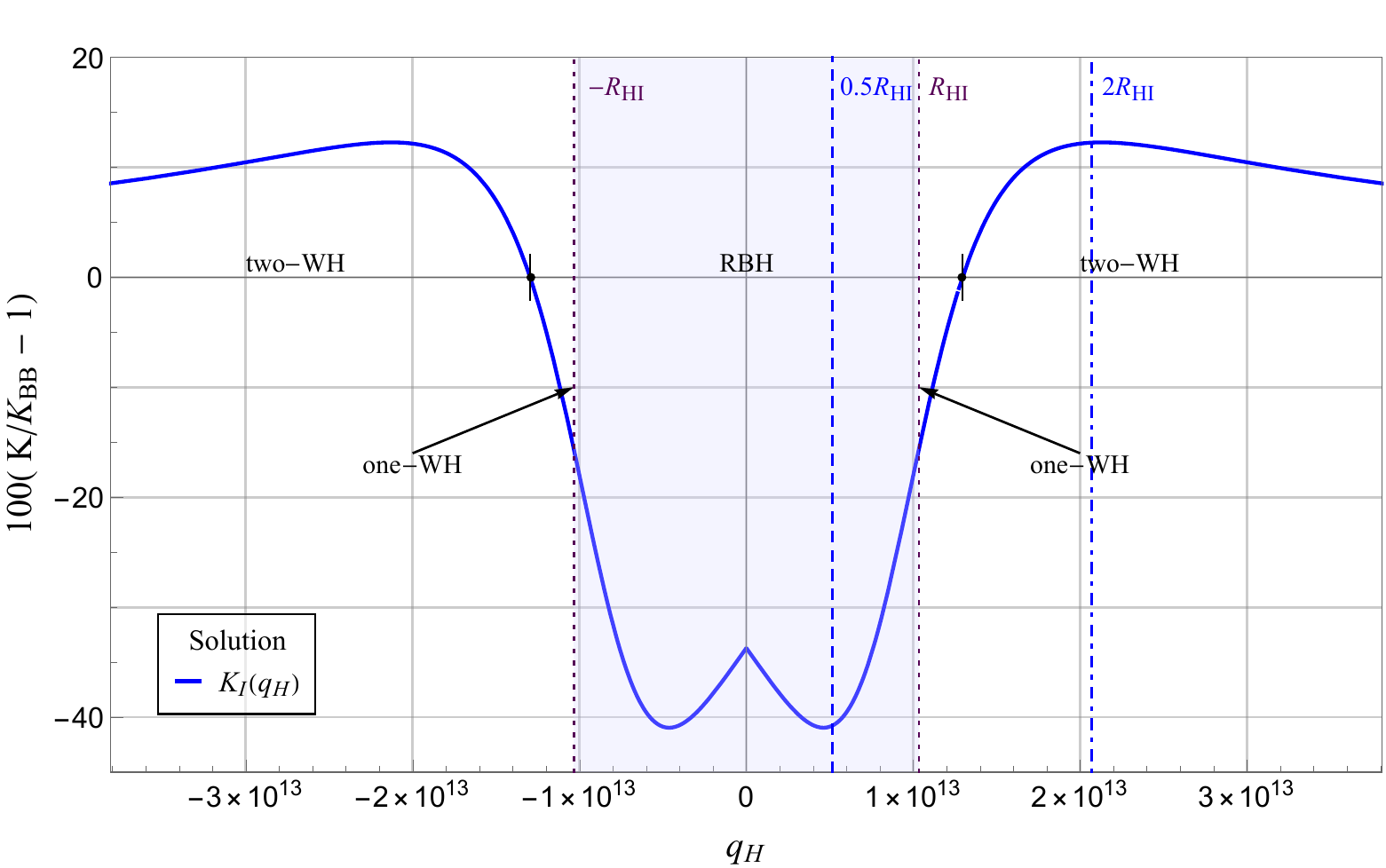}
\caption{Graphical representation of the variation of the Kretschmann scalar at $r = 0$ under changes in $q_H$, Eq.~(\ref{K_percent}), for the BB solution with a halo $K_I$ in comparison with the BB solution without a halo $K_{BB}$. Following the classifications of the BB solution with a halo, the domain of $q_H$ is divided into the following regions: $R_{HI} < q_H < -R_{HI}$ (two-WH: unshaded region); $-R_{HI} < q_H < R_{HI}$ (RBH: shaded region); and $q_H = R_{HI}$ (one-WH: dotted purple). We highlight the points corresponding to $0\%$ (black vertical dashed line) near $q_H = 0$.
} 
\label{fig:percent_K_dadosI}
\end{figure}

\begin{figure}[t!]
\includegraphics[width=\linewidth]
{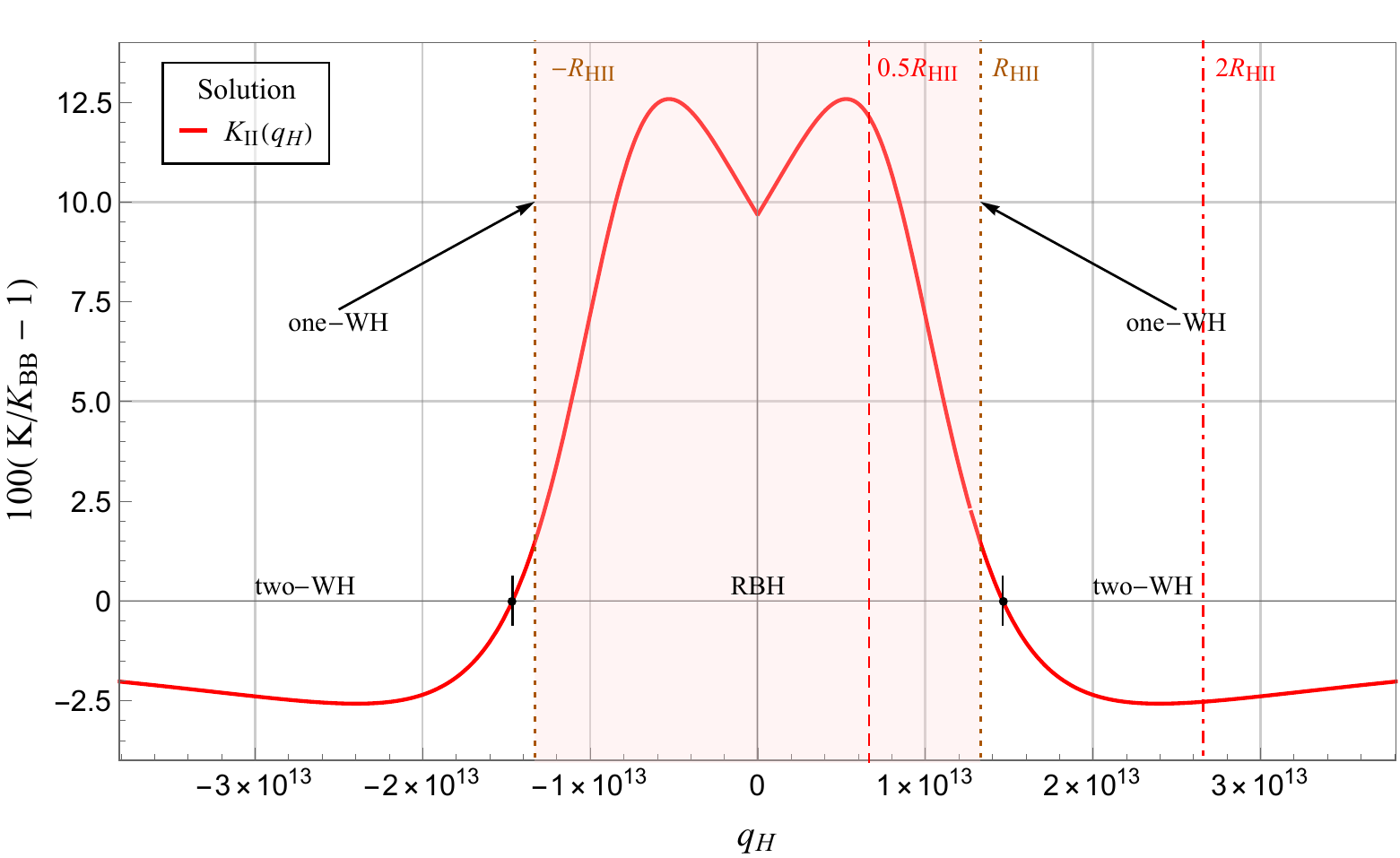}
\caption{Graphical representation of the variation of the Kretschmann scalar at $r = 0$ under changes in $q_H$, Eq.~(\ref{K_percent}), for the BB solution with a halo $K_{II}$ in comparison with the BB solution without a halo $K_{BB}$. Following the classifications of the BB solution with a halo, the domain of $q_H$ is divided into the following regions: $R_{HII} < q_H < -R_{HII}$ (two-WH: unshaded region); $-R_{HII} < q_H < R_{HII}$ (RBH: shaded region); and $q_H = R_{HII}$ (one-WH: dotted gold). We highlight the points corresponding to $0\%$ (black vertical dashed line).
} 
\label{fig:percent_K_dadosII}
\end{figure}

The  behavior of $K$ with respect to $r$ is depicted in Fig.~\ref{fig:scalarK}, which we discuss separately for the three cases underline above:

\begin{itemize}

\item The upper panel shows $K$ for $q = q_H = 0.5R$, indicating that the regularity of the BH is preserved with the inclusion of the halo in the BB solution. However, for Data I, in comparison with the BB solution, there is a decrease in the curvature as $r$ approaches the centre. This indicates that the presence of the halo with parameters $(V_{cI}, a_I)$, which implies a reduction in the BH mass $(M_I < M_0)$, results in a $38.67\%$ reduction of the curvature at the centre of the RBH, Eq.~(\ref{limit_K_r=0}), see Fig.~\ref{fig:percent_K_dadosI}.

In contrast, for Data II, there is an increase in the curvature as $r$ approaches the centre of the BH. This suggests that the presence of the halo with parameters $(V_{cII}, a_{II})$, implying an increase in the BH mass $(M_{II} > M_0)$, results in a $12.31\%$ increase in the curvature at the centre of the RBH, Eq.~(\ref{limit_K_r=0}), see Fig.~\ref{fig:percent_K_dadosII}.

\item The central panel shows $K$ for $q = q_H = R$, indicating that regularity is preserved in the presence of the DM halo. However, small variations in the curvature occur along $r$, where the curves corresponding to Data I (blue dashed) and Data II (red dot-dashed) tend to oscillate around the curve of the BB solution without a halo (purple solid).

In this regime of $q_H$, the curvature in the region of the hypersurface at $r = 0$ changes according to the adopted values of $(V_c, a)$ of the halo: for Data I, where one-WH $\to$ two-WH, there is a decrease of $1.22\%$, Eq.~(\ref{limit_K_r=0}), see Fig.~\ref{fig:percent_K_dadosI}. For Data II, where one-WH $\to$ RBH, there is an increase of $2.29\%$ in the central curvature, Eq.~(\ref{limit_K_r=0}), see Fig.~\ref{fig:percent_K_dadosII}.

In both cases, although the causal character of the hypersurface is modified, no significant changes occur in the intensity of the curvature when compared to the other scenarios in which $q_H \neq R$.

\item The lower panel shows $K$ for $q = q_H = 2R$, where the regularity of the WHs is also preserved, as well as the class (two-WH) within the Simpson–Visser solution. In this scenario, the variation of the curvature at the timelike hypersurface depends on the values of the halo parameters $(V_c, a)$: for Data I, there is an increase of $11.65\%$ in the central curvature, Eq.~(\ref{limit_K_r=0}), see Fig.~\ref{fig:percent_K_dadosI}; For Data II, there is a decrease of $2.57\%$ in the curvature, Eq.~(\ref{limit_K_r=0}), see Fig.~\ref{fig:percent_K_dadosII}.
    
\end{itemize}

Noting that the gain or loss in the intensity of the curvature at $r = 0$ also depends on the parameter $q_H$, from Eq.~(\ref{K_scalar_BB_halo}) we evaluate $K(r=0)$ for the Data I and II sets, in order to examine the behaviour of the central curvature. For this, we use the relation:
\begin{eqnarray}
    100\left( \frac{K(r=0)}{K_{BB}(r=0)} - 1\right)_{q=q_H} \ ,\label{K_percent}
\end{eqnarray}
where $K_{BB}(r)$ arises from the BB solution. For its graphical representation under Data I (Fig.~\ref{fig:percent_K_dadosI}) and Data II (Fig.~\ref{fig:percent_K_dadosII}), we again divide the domain of $q_H$ into regions according to the discussions in Subsection~\ref{subsec:event_horizon}.

\begin{itemize}

\item  (a) For $R_{HI} < q_H$ or $q_H < -R_{HI}$, corresponding to the two-WH regime, we find that for Data I, Fig.~\ref{fig:percent_K_dadosI}, only a small portion of $q_H$ indicates a decrease in the central curvature in the low-$q_H$ regime. The increase in $K$ occurs for $1.296 \times 10^{13},[m] < q_H < 3.9371 \times 10^{19},[m]$, or $1.2534 R_{HI} < q_H < 3.8076 \times 10^{6} R_{HI}$.

(b) For $R_{HII} < q_H$ or $q_H < -R_{HII}$, corresponding to the two-WH regime, we find that for Data II, Fig.~\ref{fig:percent_K_dadosII}, in addition to scenarios with both an increase (or decrease) in curvature, the points at which the increment is null lie within the classification of the bidirectional throat, where $q_H = \pm 1.4625 \times 10^{13},[m] \approx \pm 1.0996 R_{HII}$.

\item (a) For $-R_{HI} < q_H < R_{HI}$, corresponding to the RBH regime, we find that for Data I, Fig.~\ref{fig:percent_K_dadosI}, all values of $q_H$ indicate a decrease in the central curvature. The greatest mitigation of $K$ is $40.95\%$, occurring at $q_H = \pm 4.625 \times 10^{12},[m] \approx \pm 0.4473 R_{HI}$. This decrease suggests that, although the BB solution is regular by construction, the values $(V_{cI}, a_I)$ may render the regularity further removed from singular regimes. 

(b) for $-R_{HII} < q_H < R_{HII}$, corresponding to the RBH regime, we find that for Data II, Fig.~\ref{fig:percent_K_dadosII}, all values of $q_H$ indicate an increase in the central curvature. The largest increase in $K$ is $12.58\%$, corresponding to $q_H = \pm 5.287 \times 10^{12},[m] \approx \pm 0.3975 R_{HII}$. 

Although regularity is preserved, the central curvature responds to the halo parameters. If there is a decrease in the BH mass (Data I), as a response to $V_c$ and $a$, the curvature is softened. If there is an increase in the BH mass (Data II), then the curvature is enhanced. This suggests that the halo acts as a control over the internal geometry of the RBH, bringing it closer to or further from strongly curved regimes.

\item (a) For $q_H = \pm R_{HI}$, corresponding to the one-WH regime, we find that for Data I, Fig.~\ref{fig:percent_K_dadosI}, no discontinuity appears in the curve of the function, indicating that, from the perspective of the scalar curvature, the transition RBH $\to$ one-WH occurs seamlessly, carrying a decrease of $15.62\%$ under the presence of the halo;

(b) For $q_H = \pm R_{HII}$, corresponding to the one-WH regime, we find that for Data II, Fig.~\ref{fig:percent_K_dadosII}, the curve of the function is also continuous and finite, maintaining a continuous behaviour from the perspective of the scalar curvature, with an increase of $1.46\%$ under the presence of the halo.

In both cases, the value of the central curvature remains finite and continuous at $q_H = \pm R_H$, indicating that the change in the causal character of the hypersurface (timelike $\to$ null) does not imply a breakdown of the geometry, that is, it corresponds to a continuous transition from a geometrical standpoint in the presence of the DM halo.
    
\end{itemize}

\subsection{Black Hole Shadow Radius}\label{subsec:shadow}

A prominent observational channel to peer into the nature of BHs comes from the extreme deflection of the trajectories of photons as they pass through the vicinity of the photon sphere -- a spherical region around the BH in which these particles can orbit in unstable circular trajectories --. This feature of the space-time allows a distant observer, located at radial coordinate $r_O$, to image the BH in terms of a bright ring of radiation surrounding a central dark region, canonically interpreted as the BH shadow \cite{shadow}. For the sake of this section we use the formalism presented in \cite{shadow,Perlick:2021aok} to determine the radius of such a dark region (i.e. shadow radius), denoted by $r_{sh}$, for the space-time metric (\ref{metrica_geral_BB}). To find it, let us first define the auxiliary function $h(r)$ as
\begin{equation}
	h(r) = \sqrt{\frac{\Sigma(r)^2}{A(r)}} \ ,
\end{equation}
such that the radius of the photon sphere, denoted by $r_{ph}$, is obtained via the equation
\begin{equation}
	\left. \frac{d h^2(r)}{dr} \right|_{r = r_{ph}} = 0 \,.
\end{equation}
which reads explicitly as the implicit equation
\begin{equation}
    \left. \left[\frac{d\Sigma(r)}{dr}A(r)-\frac{\Sigma(r)}{2}\frac{dA(r)}{dr}\right] \right|_{r = r_{ph}}=0 \ .\label{cond.photon.sphere}
\end{equation}
Upon solving of the above equation for $r_{ps}$, the shadow radius for an observer at radial position $r_O$ is defined by
\begin{equation}
	r_{sh} = r_{ph} \sqrt{ \frac{A(r_O)}{A(r_{ph})} } \ ,\label{r_sh_GERAL_naoplana}
\end{equation}
where $A(r_O)$ denotes the value of the metric function at the observer's location $r_O$.

\begin{figure}[t!]
\includegraphics[width=\linewidth]
{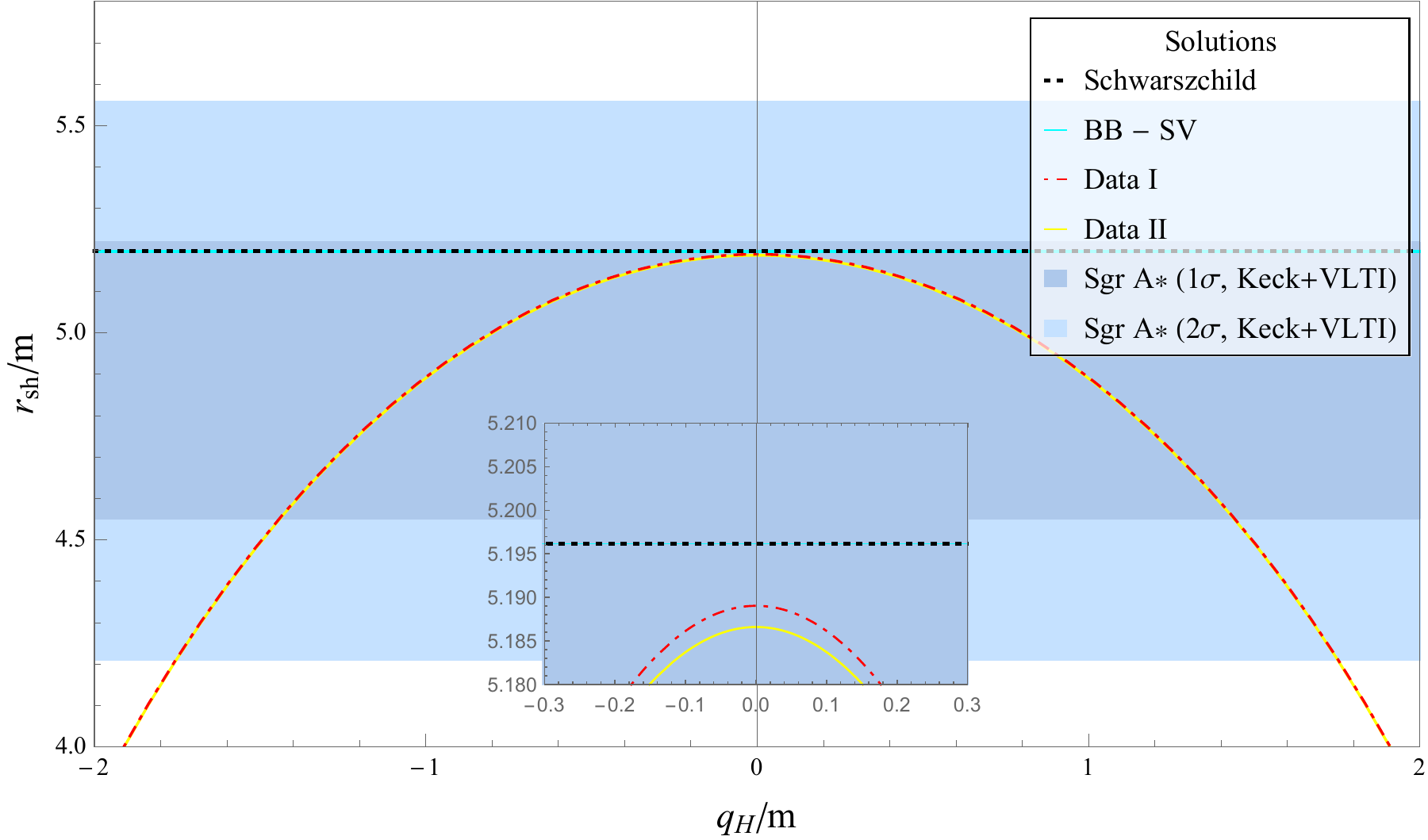}
\caption{Graphical representation of the shadow radius $r_{sh}$ under some values of the parameter $q_H$ for the BB solution with a halo — for values of Data I and Data II — in comparison with the BB solution without a halo (Simpson–Visser BB), and additionally with the Schwarzschild solution without a halo. The blue shaded and natural regions correspond to the allowed regions for the shadow radius according to the EHT constraints at $1\sigma$ and $2\sigma$, respectively.} 
\label{fig:rsh_q}
\end{figure}

Substituting first the temporal metric coefficient from Eq.~(\ref{sol_halo_BB}) into the condition for the photon sphere $r_{ph}$ in Eq.~(\ref{cond.photon.sphere}), we then compute the shadow radius $r_{sh}$ from Eq.~(\ref{r_sh_GERAL_naoplana}). Fig.~\ref{fig:rsh_q} shows the behaviour of $r_{sh}$ under variations of $q_H$, in order to define a valid domain for $q_H$ based on the intervals (shaded regions) inferred from observational data of Sgr A* obtained by the EHT. These regions represent the statistical limits of $1\sigma$, with a $68\%$ confidence level, and $2\sigma$, with a $95\%$ confidence level, for $r_{ph}$. For the values of Data I and II, a significant change in the behaviour of the shadow is observed, decreasing as $|q_H|$ increases. However, the $1\sigma$ and $2\sigma$ regions suggest limiting values for $q_H$, given by
\begin{equation}
    -1.7538\leq q_H/M_I\leq1.7538 \ ,
\end{equation}
for Data I, and
\begin{equation}
    -1.7522\leq q_H/M_{II}\leq1.7522 \ ,
\end{equation}
for Data II. The value $q_H = 0$ corresponds to the Schwarzschild-type solution with a DM halo. For both datasets, it follows that $q_H/m \leq R_H/m$, suggesting that the intervals of $q_H/m$ lie within the domain in which $A(r)$ admits two real horizons — in the presence of the halo, the value of $q_H$ implies an RBH-type structure. This suggests that, under the analogy between the parameters of M60 and the observational data of Sgr A*, the BB solution proposed here does not allow for the existence of a WH structure immersed in a DM halo.

\subsection{Effective Potential for Massless Particles}\label{subsec:Veff_massless}

Starting from the line element that describes the BB spacetime, Eq.~(\ref{metrica_geral_BB}) together with Eq.~(\ref{sol_halo_BB}), the trajectory of particles is governed by the Lagrangian ${\mathcal L}$.
\begin{equation}
    {\mathcal L}=\dot{s}^2=A(r)\dot{t}^2-\frac{\dot{r}^2}{A(r)}-\Sigma(r)^2\left(\dot{\theta}^2+\sin^2{\theta}\dot{\phi}^2\right) \ ,
\end{equation}
where the dot denotes differentiation with respect to the affine parameter $\tau$ \cite{Silva:2024fpn}. Using the Euler–Lagrange equations
\begin{equation}
    \frac{d}{d\tau}\left(\frac{\partial{\mathcal L}}{\partial\dot{x}^{\mu}}\right)-\frac{\partial{\mathcal L}}{\partial x^{\mu}}=0\ ,
\end{equation}
and considering the spherically symmetric character of the spacetime, for which we assume $\theta = \pi/2$ without loss of generality. Thus, we obtain the following equations of motion for massless particles
\begin{eqnarray}
    A(r)\dot{t}&=&E \ , \label{eq.mov01} \\
    A(r)\dot{t}^2-\frac{\dot{r}^2}{A(r)}-\Sigma(r)^2\dot{\phi}^2&=&0 \ , \label{eq.mov02} \\
\Sigma(r)^2\dot{\phi}&=&l \ ,
\end{eqnarray}
where $E$ and $l$ are the energy and angular momentum of the particle, respectively. It then follows that
\begin{equation}
E^2=\dot{r}^2+\frac{A(r)l^2}{\Sigma(r)^2} \ ,
\end{equation}
which can be rearranged as
\begin{equation}
\dot{r}^2=E^2-V_{eff}\ , \label{E_Veff}
\end{equation}
where $V_{eff}$ is the effective potential given by
\begin{equation}
V_{eff}=l^2\frac{A(r)}{\Sigma(r)^2} \ .\label{Veff}
\end{equation}
The orbits may be either stable or unstable. If they are unstable, the particles exhibit a radial acceleration $\alpha = \ddot{r}$, which can be obtained as
\begin{equation}
    \alpha=-\frac{1}{2}\frac{dV_{eff}}{dr} \ .\label{aceleracao}
\end{equation}

We obtain the effective potential for the BB model with a DM halo by substituting Eq.~(\ref{sol_halo_BB}) into Eq.~(\ref{Veff}), yielding
\begin{equation}
V_{eff}=l^2\left[\frac{\Sigma(r)-2m}{\Sigma(r)^{3}}-
    \frac{2V_c^2}{a^2+\Sigma(r)^2}\right] \ .\label{Veff_halo}
\end{equation}
Thus, from Eq.~(\ref{aceleracao}), we obtain the radial acceleration
\begin{equation}
\alpha(r)=l^2r\left[\frac{\Sigma(r)-3m}{\Sigma(r)^{5}}-\frac{2V_c^2}{(a^2+\Sigma(r)^2)^2}
 \right]  \ .\label{aceleracao_halo}
\end{equation}
Note that for Eqs.~(\ref{Veff_halo}) and (\ref{aceleracao_halo}), if $V_c = 0$, we recover $V_{eff}$ and $\alpha(r)$ for the BB solution without a halo \cite{Silva:2024fpn}.

\begin{figure}[htbp]
    \centering
    \begin{subfigure}
        \centering
        \includegraphics[width=\linewidth]{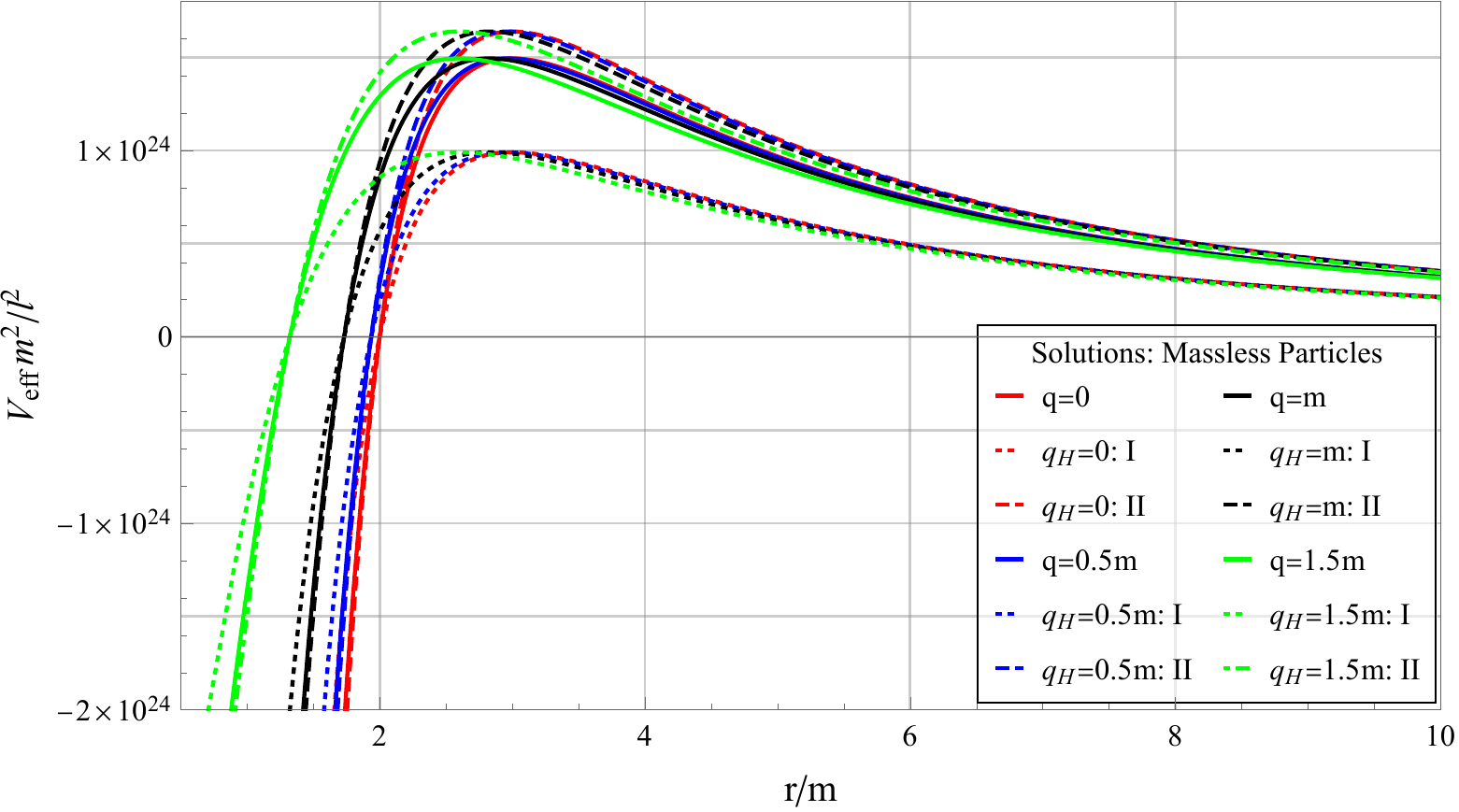}
    \end{subfigure}

    \begin{subfigure}
        \centering
        \includegraphics[width=\linewidth]{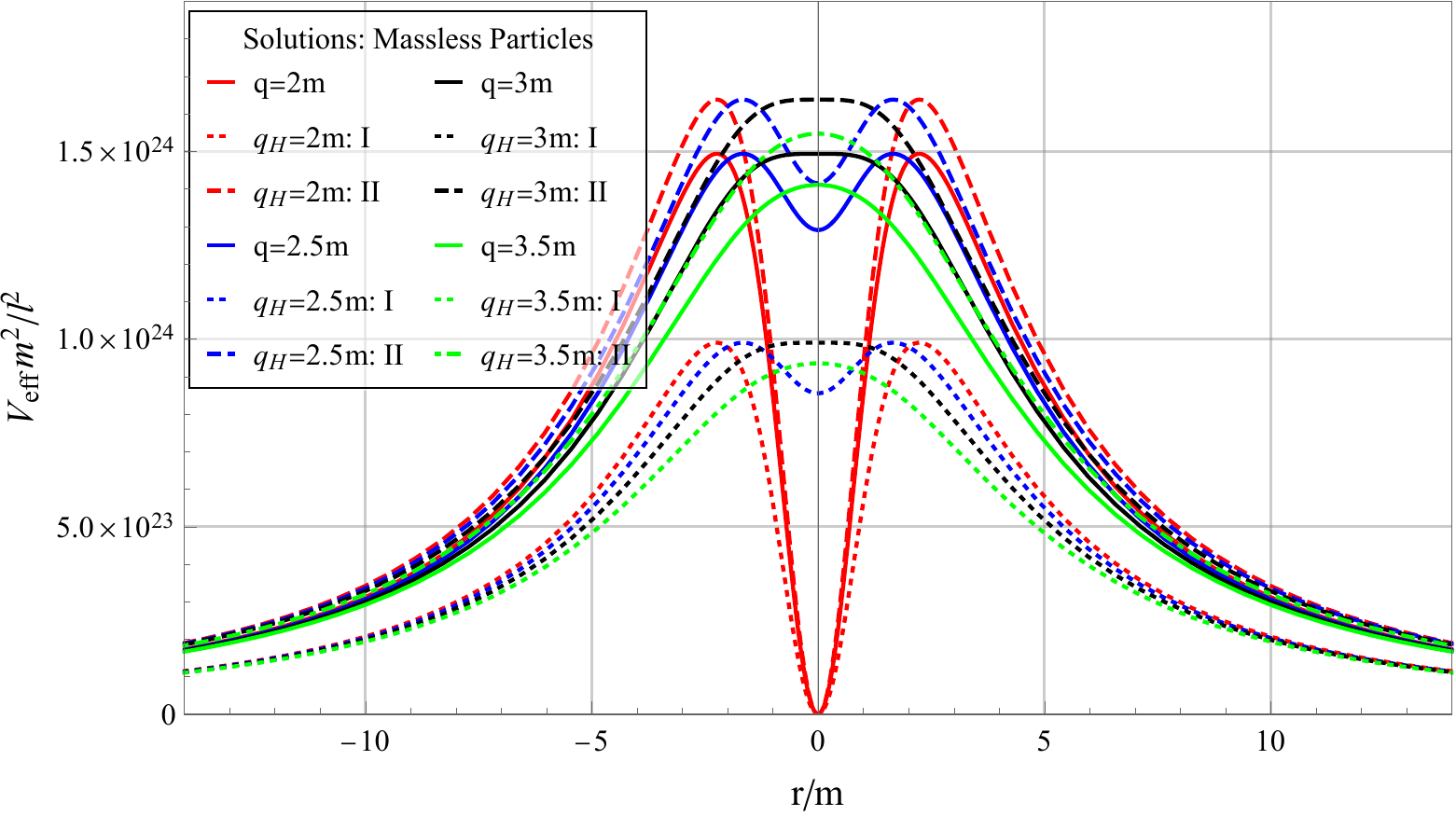}
    \end{subfigure}
    \caption{Graphical representation of the effective potential $V_{eff}(r)$ of the BB solution with a halo, from Eq.~(\ref{Veff_halo}), for massless particles — for values of Data I and Data II — with different values of the parameter $q_H$, following null geodesics of the metric given in Eq.~(\ref{metrica_geral_BB}). This also includes the BB solution without a halo. The upper panel corresponds to values $q_H < 2m$, while the lower panel corresponds to values $q_H \geq 2m$.
}
    \label{fig:Veff_massless}
\end{figure}

Fig.~\ref{fig:Veff_massless} shows the behaviour of the effective potential from Eq.~(\ref{Veff_halo}) for the BB solution with and without a halo. For $q_H < 2m$ (upper panel), a single maximum $r_{1} > 0$ is preserved in the presence of the halo, representing the radius of an unstable circular orbit of a massless particle. For $2m \leq q_H \leq 3m$ (lower panel), two maxima $r_1 > 0$ and $r_2 < 0$, and one minimum $r_3 = 0$ are maintained with the halo, indicating, respectively, two unstable orbits and one stable orbit. For $q_H > 3m$ (lower panel), a single unstable orbit at $r_3 = 0$ is preserved.

The correction to the potential due to the halo is manifested through vertical shifts of the maxima and minima. At these points, the radial velocity vanishes, $\dot{r} = 0$, which, from Eq.~(\ref{E_Veff}), implies that $(E^2=V_{eff})_{r_{1,2,3}}$. Therefore: for Data I, less energy is required for the particle to escape, $V_{eff} < V_{eff,BB} \to E^2 < E^2_{BB}$; in contrast, for Data II, more energy is required for the escape of the massless particle, $V_{eff} > V_{eff,BB} \to E^2 > E^2_{BB}$.

When a particle, coming from infinity with energy $E_p$, reaches an extremum of the potential $V_{eff}$, where $\dot{r} = 0$, its angular velocity $\dot{\phi}$ can be obtained from Eqs.~(\ref{eq.mov01})~and~(\ref{eq.mov02}), given by
\begin{equation}
    \frac{E_p^2}{\Sigma(r_{1,2,3})^2A(r_{1,2,3})}=\dot{\phi}^2 \ .
\end{equation}
The circular character of the unstable orbit is preserved. We verify modifications in the angular velocity $\dot{\phi}$ when comparing the cases with halo, $\dot{\phi}_{I,II}$, and without halo, $\dot{\phi}_{BB}$, for which, in the regime $0 \leq q_H < 3m$, we have that
\begin{equation}
    \frac{\dot{\phi}^2_I}{\dot{\phi}^2_{BB}}\approx1.5086  \ , \label{razao_vel_ang_I}
\end{equation}
for Data I, and
\begin{equation}
    \frac{\dot{\phi}^2_{II}}{\dot{\phi}^2_{BB}}\approx0.9118  \ , \label{razao_vel_ang_II}
\end{equation}
for Data II. 

Eqs.~(\ref{razao_vel_ang_I}) and (\ref{razao_vel_ang_II}) show that the angular velocity $\dot{\phi}$ responds differently for each dataset considered here. In order to verify this, we separate the asymptotic behaviour of $\dot{\phi} = \dot{\phi}(V_c, a)$ into two domains, given that $(M_I, V_{cI}, a_I) < (M_{II}, V_{cII}, a_{II})$: the first corresponds to Data I, when $(V_c, a)$ are small
\begin{equation}
\dot{\phi}_I=\dot{\phi}_{BB}+
\left(1-\frac{a^2}{\Sigma(r_{1})^2}\right)\frac{2V_c^2}{\left(\Sigma(r_{1})^2-2m\right)^2} + {\mathcal O(V_c^3,a^3)}\ ,\label{aprox_vel_ang_I}
\end{equation}
and the second corresponds to Data II, when $(V_c, a)$ are large
\begin{eqnarray}
	&&\dot{\phi}_{II}=\dot{\phi}_{BB}+\frac{2\Sigma(r_1)^2V_c^2}{a^2\left(\Sigma(r_1)^2-2m\right)^2}  
		\nonumber \\
	&& \times\Bigg[\frac{\Sigma(r_1)^2}{a^2}\Bigg(\frac{2\Sigma(r_1)}{\Sigma(r_1)-2m}
	-1\Bigg)+1 \Bigg]\nonumber \\
	&&+a^{-4}{\mathcal O\left(V_c^{-5},a^{-1}\right)} \ .\label{aprox_vel_ang_II}
\end{eqnarray}

For the first scenario of Eq.~(\ref{aprox_vel_ang_I}), if $V_c \to 0$ we recover $\dot{\phi}_{BB}$. For low velocities, $V_c \ll 1$, the effect of the halo can be treated as a perturbation. The increase in the angular velocity in Eq.~(\ref{razao_vel_ang_I}) can be justified by the scale of the halo radius $a$ and the low intensity of its critical velocity $V_c$ -- which is indeed small -- in \cite{mod1}. If $\Sigma(r_1)^2 \gg a^2$, the correction depends only on $V_c$. On the other hand, Eq.~(\ref{razao_vel_ang_I}) suggests the condition $\Sigma(r_1)^2 > a^2$. This indicates that the BB radius $\Sigma(r_1)$ is located in a region where the influence of the halo on the local geometry is weaker. Thus, a positive correction to the angular velocity occurs.

For the second scenario of Eq.~(\ref{aprox_vel_ang_II}), there is no contribution if the halo is very diffuse, that is, if $a \to \infty$, we recover $\dot{\phi}_{BB}$. The decrease in the angular velocity in Eq.~(\ref{razao_vel_ang_II}) can be justified by the large extent of the halo radius $a$ and by higher values of $V_c$ -- which are still observationally constrained to $V_c \ll 1$ \cite{mod1}. Note also that Eq.~(\ref{razao_vel_ang_II}) suggests a condition on the mass as given by Eq.~(\ref{aprox_vel_ang_II}),
\begin{equation}
    1<\frac{2m}{\Sigma}<\frac{a^2+\Sigma^2}{a^2-\Sigma^2} \ ,
\end{equation}
such that $a > \Sigma$, indicating that the halo is located externally to the characteristic radius of the BB.

\begin{figure}[t!]
    \centering
    \begin{subfigure}
        \centering
        \includegraphics[width=\linewidth]{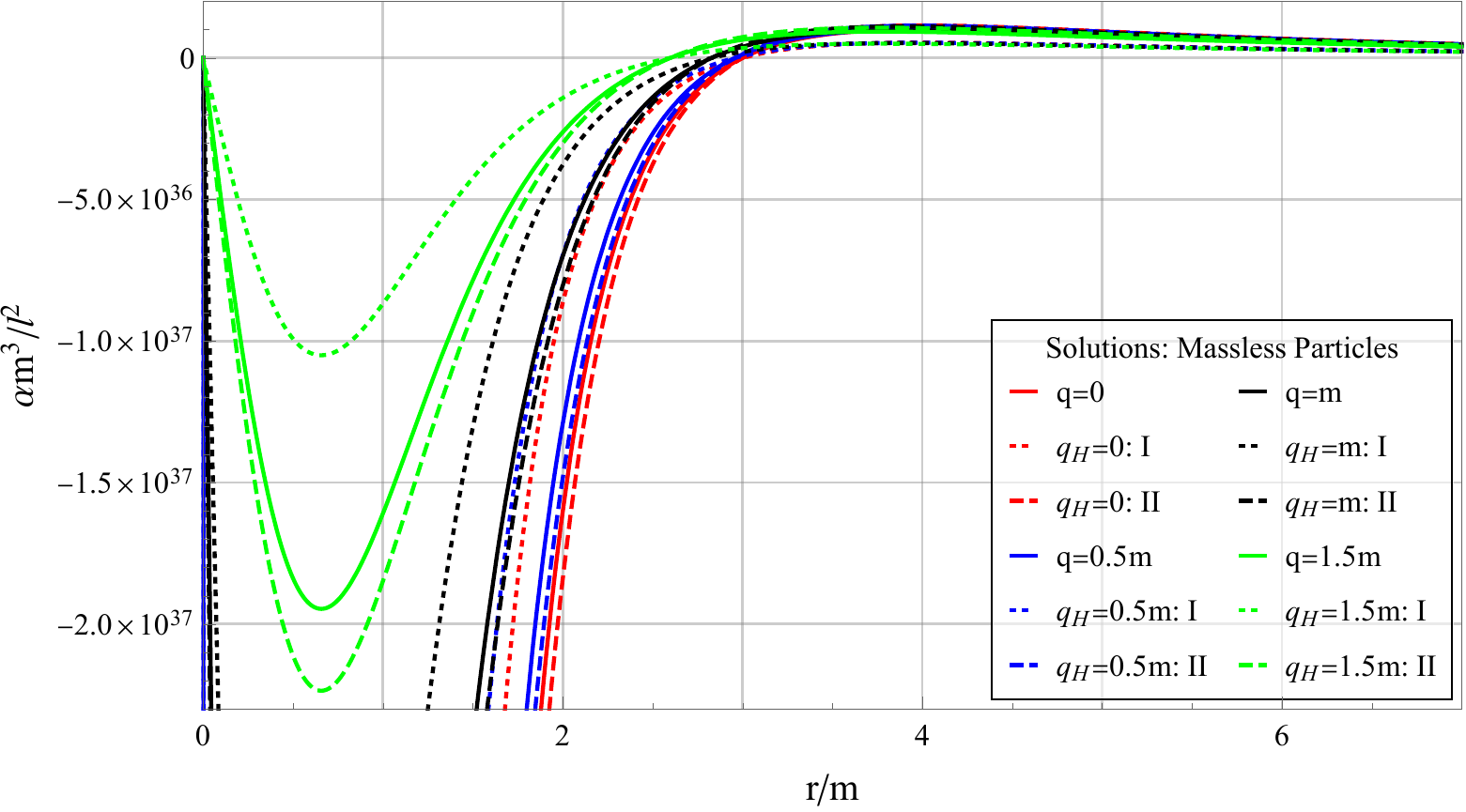}
    \end{subfigure}

    \begin{subfigure}
        \centering
        \includegraphics[width=\linewidth]{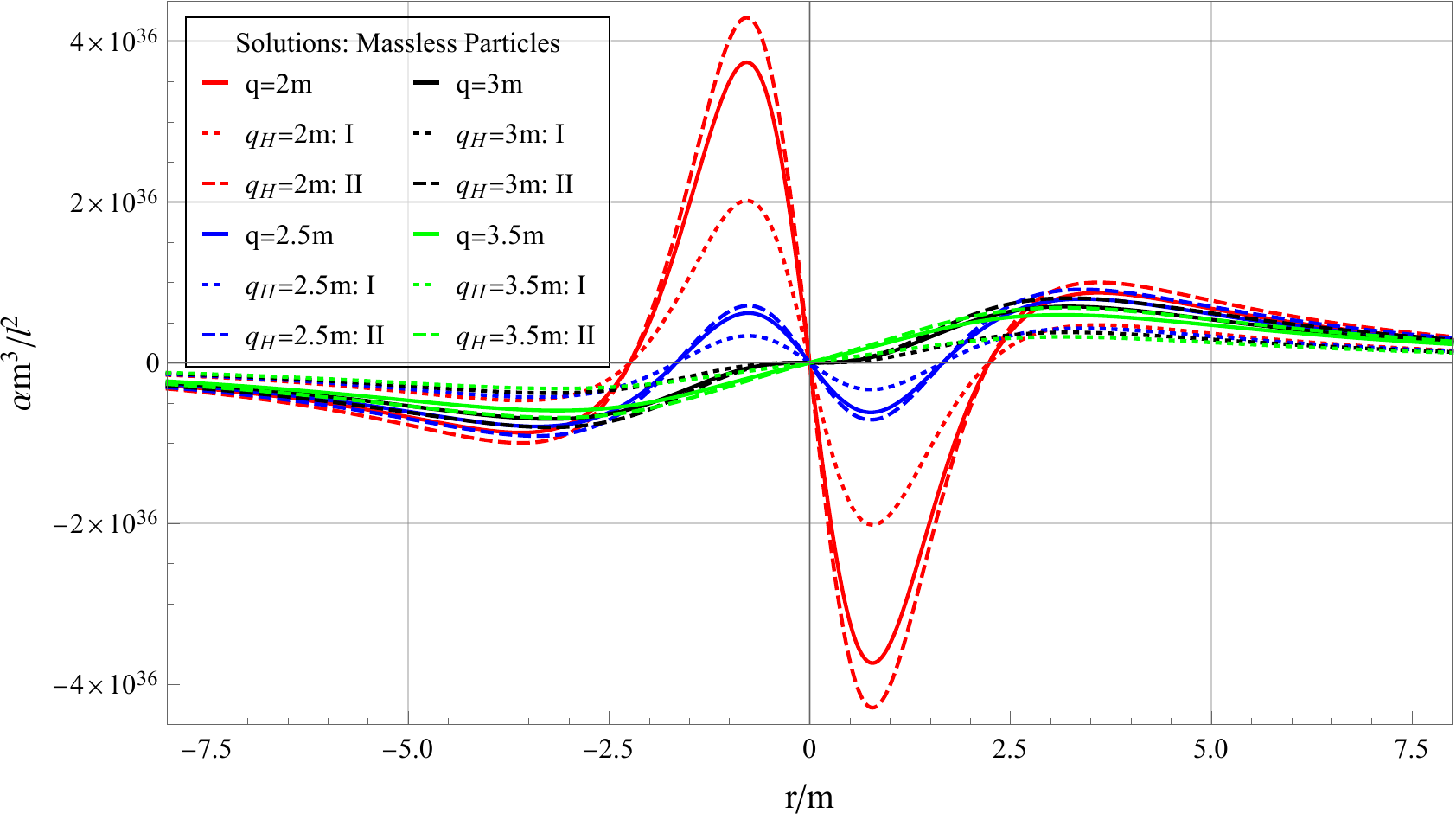}
    \end{subfigure}
    \caption{Graphical representation of the radial acceleration $\alpha(r)$ of the BB solution with a halo, from Eq.~(\ref{aceleracao_halo}), for massless particles — for values of Data I and Data II — with different values of the parameter $q_H$, following null geodesics of the metric given in Eq.~(\ref{metrica_geral_BB}). This also includes the BB solution without a halo. The upper panel corresponds to values $q_H < 2m$, while the lower panel corresponds to values $q_H \geq 2m$.
}
    \label{fig:alpha_massless}
\end{figure}

For completeness of our analysis, Fig.~\ref{fig:alpha_massless} shows the radial acceleration $\alpha(r)$ from Eq.~(\ref{aceleracao_halo}) for a massless particle in the BB solution with a halo and without a halo, $\alpha(r)|_{(V_c=0,a=0)}=\alpha_{BB}(r)$. In the regime $0 \leq q_H \leq 3m$, the acceleration remains positive (repulsive) when $r > r_1$ and negative (attractive) when $r < r_1$. However, the contribution of the halo for the datasets considered here shows an inversion of behaviour taking $r = r_1$ as the transition point: for Data I, we have that $(\alpha_I<\alpha_{BB})_{r>r_1}$, and $(\alpha_I>\alpha_{BB})_{r<r_1}$; in contrast, for Data II, we have that
$(\alpha_{II}>\alpha_{BB})_{r>r_1}$, and $(\alpha_{II}<\alpha_{BB})_{r<r_1}$.

The largest variation in the intensity of the acceleration occurs in the region $r < r_1$ for $0 \leq q_H < 3m$, indicating that the halo enhances the attraction of the particle when there is an increase in the BH mass and in the halo velocity $V_c$, as observed for Data II. However, this acceleration can be mitigated when the BH mass and the velocity $V_c$ are reduced, as observed for Data I. We emphasize that, as discussed in \cite{mod1}, the velocity $V_c$ increases (or decreases) as the radius $a$ also increases (or decreases), indicating that the influence of the halo on the acceleration $\alpha(r)$ is bounded by an upper limit on the radius $a$ and a lower limit on the velocity $V_c$ — from Eq.~(\ref{aceleracao_halo}), if $V_c \to 0$ or $a \to \infty$, then $\alpha(r) \to \alpha_{BB}(r)$.

Similarly, when $q_H > 3m$, the intensity of the particle’s repulsive acceleration is reduced for Data I, but enhanced for Data II.

\subsection{Thermodynamic Properties}\label{subsec:Thermodynamic properties}

\begin{figure}[htb!]
\includegraphics[width=\linewidth]
{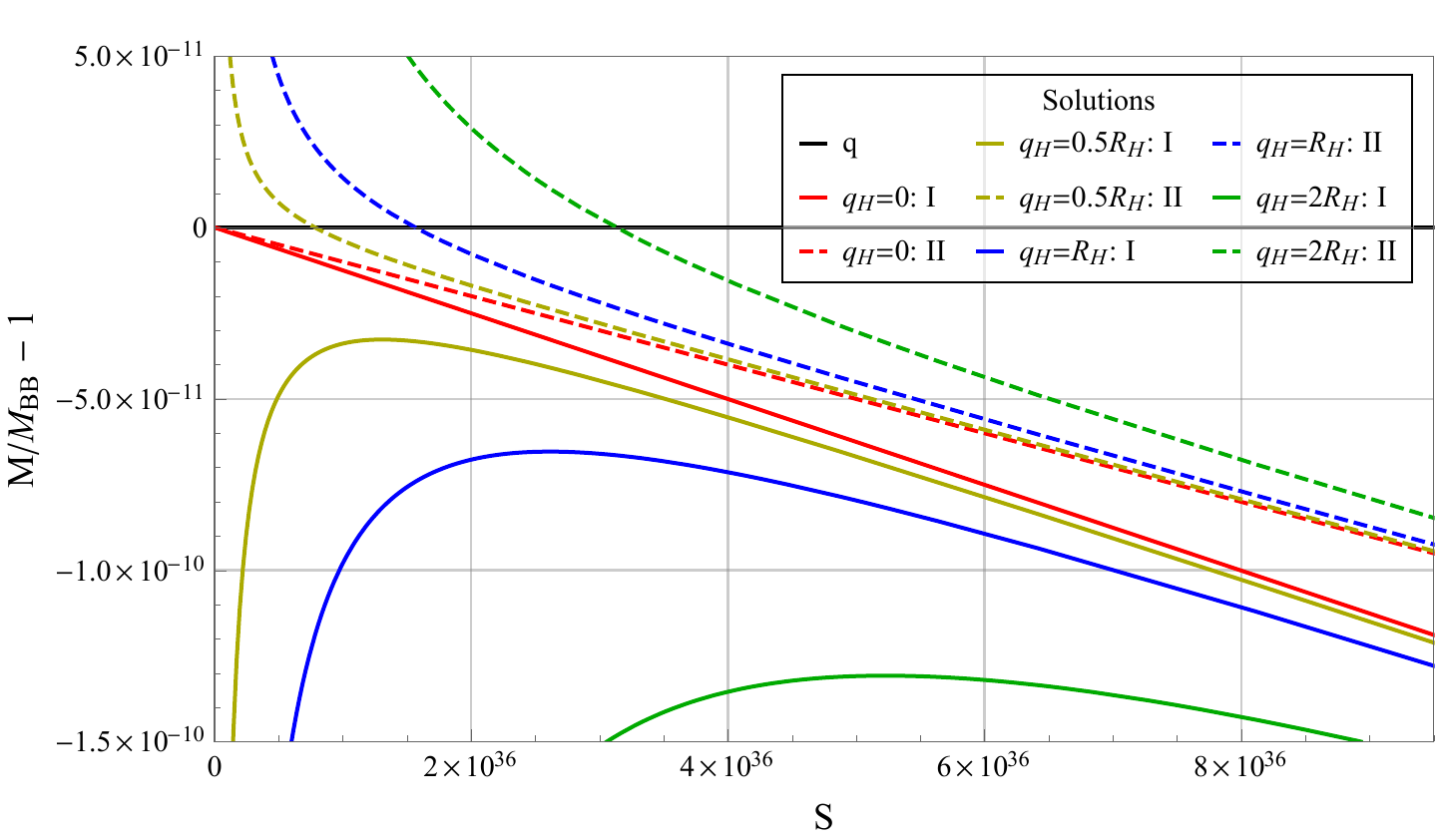}
\caption{Graphical representation of the BH mass-entropy relation, $M/M_{BB} - 1$, under some values of the parameter $q_H$ for the BB solution with a halo — for values of Data I and Data II — in comparison with the BB solution without a halo. The values of $q_H$ are chosen in terms of the event horizon radius of the Schwarzschild-type solution with a halo, $R_H$.
}
\label{fig:massa_entropia_BB_halo}
\end{figure}

Following the Bekenstein-Hawking formulation \cite{Hawking:1975vcx}, the entropy $S$ of a BH can be defined in terms of the area $\mathcal{A} = 4\pi r_h^2$, where $r_h$ is the radius of the event horizon, given by $S = \mathcal{A}/4$, such that
\begin{equation}
    S=\pi r^2_h\, .\label{S_definicao}
\end{equation}
Given the relation in Eq.~(\ref{S_definicao}), we substitute the radius $r_h$ into $F(r_h) = 0$ from Eq.~(\ref{sol_halo_BB}), obtaining the following relation for the BH mass
\begin{align}
    &M\left(S,V_c,a,q_H\right)=\sqrt{q^2_H+\frac{S}{\pi}}
    \nonumber\\
    &\times \frac{a^2\pi-\left(S+\pi q^2_H\right)\left(2V_c^2-1\right)}{2\left[S+\pi\left(a^2+q^2_H\right)\right]} \ ,\label{f_massa_entropia}
\end{align}
which recovers the relation for the BB case without a halo: $M\left(S,V_c,a,q_H\right)|_{V_c=0,a=0}=M_{BB}\left(S,q\right)$. Similarly, the Schwarzschild case with a halo is recovered when $q_H = 0$ \cite{Lobo2025H1}. Fig.~\ref{fig:massa_entropia_BB_halo} illustrates the behaviour of the mass ratio $M/M_{BB} - 1$ as a function of the entropy $S$ for BB-type BH solutions with a DM halo. A general behaviour is preserved for large values of entropy under the values of the $q_H$, given by
\begin{equation}
\lim_{S\to\infty}\left(M_{I,II}/M_{BB}\right)\to1-2V_c^2\label{lim_S_alto}\ ,
\end{equation}
indicating that, in this regime, only the parameter $V_c$ contributes to the change in the mass behaviour due to the halo.

For small values of entropy, the behaviour exhibits a more explicit dependence on the radius $a$ and the velocity $V_c$,
\begin{equation}
    \frac{M_{I,II}}{M_{BB}}= \frac{q_H\left[1+\left(\frac{q_H}{a}\right)^2\left(1-2V_c^2\right)\right]}{q\left[1+\left(\frac{q_H}{a}\right)^2\right]}+{\mathcal{O}(S)} \ ,\label{razao_massa_aprox}
\end{equation}
which shows that the halo ceases to have an influence if it is sufficiently far from the BH centre, i.e., when $a \gg q_H$. The same behaviour is recovered for vanishing halo rotation, $V_c = 0$, or if it is low enough $V_c \ll V_{cI,II}$.

Fig.~\ref{fig:massa_entropia_BB_halo} shows that the mass responds differently to the entropy for Data I and II. We verify this by considering the interval $q_H \sim r_{h,s}$, relating it to the RBH, one-WH, and two-WH classes.

 \begin{itemize}

\item For Data I: except for $q_H = 0$, there is no value of $S$ for which the correspondence $M \to M_{BB}$ occurs, maintaining a suppressed mass response due to the relation $R_{HI} > r_{hI} \to q > q_H$ in the regime $S \ll 1$, as can be seen from Eq.~(\ref{razao_massa_aprox}),
\begin{equation}
    \left(1 +\frac{q_H^2}{a^2}\right) \left(1 -\frac{q}{q_H} \right) < 2 V_c^2 \frac{q_H^2}{a^2} \ ,
\end{equation}
and due to the nonzero velocity $V_c$ when $S \gg 1$, see Eq.~(\ref{lim_S_alto}).

\item For Data II: there exists a value of entropy $S_{0,II}$ such that $(M=M_{BB})|_{S_{0,II}}$, which divides the mass behaviour into two regimes. If $S < S_{0,II}$, the mass response is enhanced, $M > M_{BB}$. If $S > S_{0,II}$, the mass response is attenuated with respect to entropy, $M < M_{BB}$. If $S = S_{0,II}$, no modification to the mass is observed, even though $q_H \neq 0$, suggesting that the halo does not influence the system at this point. When $q_H = 0$, then $S_{0,II} = 0$.

\end{itemize}

\begin{figure}[t!]
\includegraphics[width=\linewidth]
{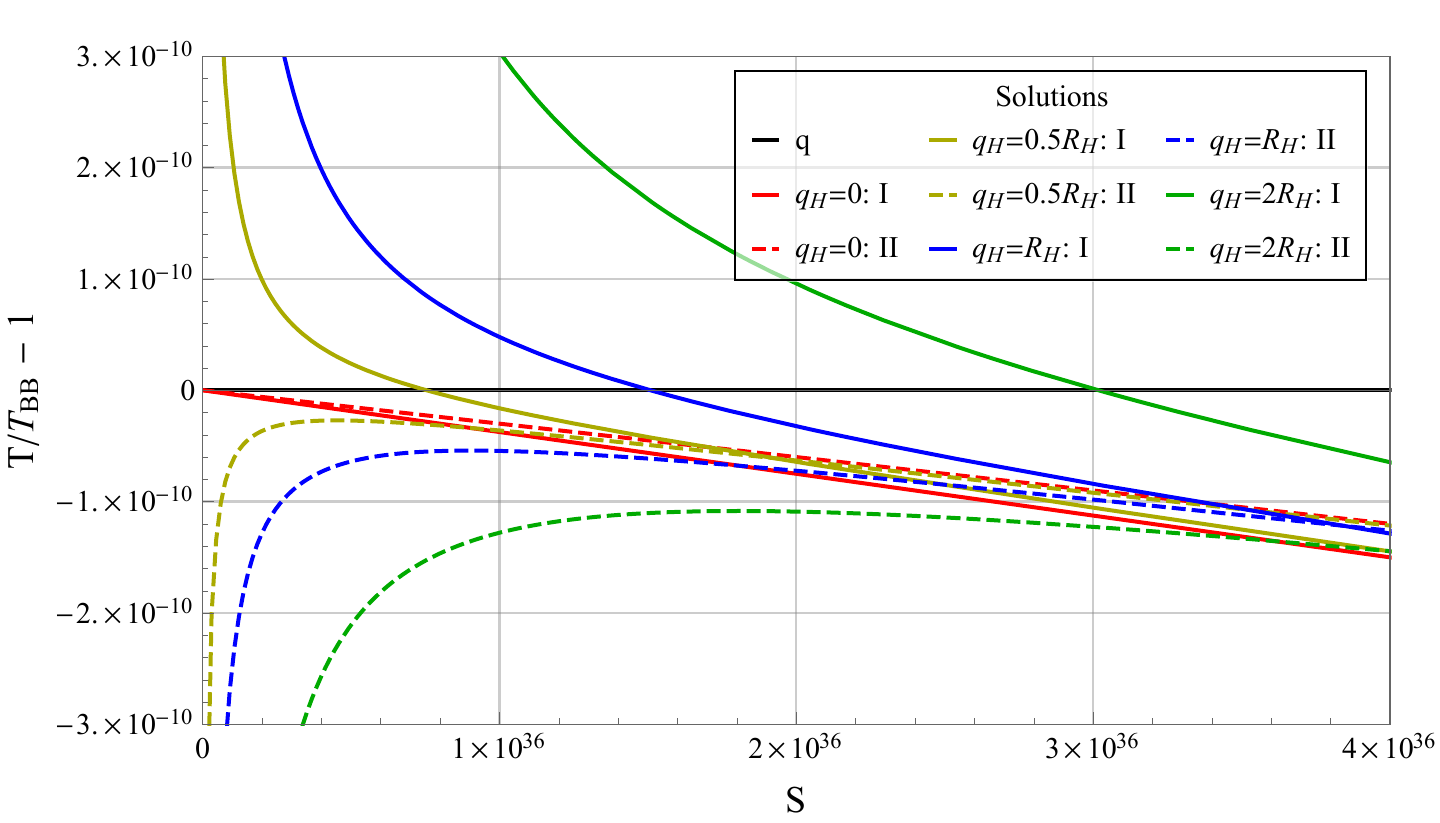}
\caption{Graphical representation of the temperature-entropy relation, $T/T_{BB}-1$, under some values of the parameter $q_H$ for the BB solution with halo — for Data I and Data II — compared with the BB solution without halo. The values of $q_H$ are chosen relative to the horizon radius of the Schwarzschild-type solution with halo, $R_{H}$.
}
\label{fig:temperatura_entropia_BB_halo}
\end{figure}

For the calculation of the temperature, we use $T = \frac{\partial M}{\partial S}$, from Eq.~(\ref{f_massa_entropia}). Thus,
\begin{equation}
T(S,V_c,a,q_H)=\frac{(\pi a^2+S_{q_H})^2-2V^2_cS_{q_H}(S_{q_H}-3\pi a^2)}{4\sqrt{\pi S_{q_H}}(\pi a^2+S_{q_H})^2}\ ,
\end{equation}
where $S_{q_H} = S + \pi q_H^2$, so that we recover the temperature in the BB case without a halo when the velocity is set to zero, $T|_{V_c=0}\to T_{BB}$. Fig.~\ref{fig:temperatura_entropia_BB_halo} shows the behaviour of the temperature ratio $T/T_{BB} - 1$ as a function of the entropy $S$ for BB-type BH solutions with a DM halo. A general behaviour is preserved for large values of entropy, given by
\begin{equation}
    \lim_{S\to\infty}\left(T_{I,II}/T_{BB}\right)\to1-2V_c \ ,
\end{equation}
indicating that, in this regime, only the velocity influences the variation of the surface temperature. For small values of entropy, the behaviour more clearly shows its dependence on the velocity $V_c$ and the radius $a$,
\begin{equation}
    \frac{T_{I,II}}{T_{BB}}=\frac{q\left[\left(1+\frac{q_H^2}{a^2}\right)^2-2V_c\frac{q_H^2}{a^2}\left(3+\frac{q_H^2}{a^2}\right)\right]}{q_H\left(1+\frac{q_H^2}{a^2}\right)^2}+{\mathcal O}(S)\ ,
\end{equation}
for which a correspondence is recovered if the halo is sufficiently far away, $a \gg q_H$, or if the halo velocity is very low, $V_c \ll V_{cI,II}$.

When we define the values of Data I and II, different temperature behaviours are observed between these two sets, indicating that small variations in $(V_c, a)$ have an influence on the thermodynamics of the BB. As shown in Fig.~\ref{fig:temperatura_entropia_BB_halo}, we select values $q_H \sim R$ in order to examine the temperature within the RBH, one-WH, and two-WH domains.

\begin{itemize}

\item For Data I: there exists a value of entropy $S_{0,I}$ that divides the overall behaviour of $T$ into two regimes. If $S < S_{0,I}$, the temperature increases; if $S > S_{0,I}$, it decreases; and if $S = S_{0,I}$, then $T \to T_{BB}$, indicating the absence of halo influence. When $q_H = 0$, we have $S_{0,I} = 0$.

\item For Data II: there is only a decrease in temperature with respect to entropy, and this decrease becomes more pronounced as $q_H$ increases. This suggests that the RBH class may exhibit the smallest halo influence on its temperature, in contrast to the two-WH class, which can exhibit the largest influence. There is no value of entropy, for $q_H \neq 0$, at which $T \to T_{BB}$.

\end{itemize}

As in the case of the mass $M(S)$, the temperature $T(S)$ is strongly affected by the parameter $q_H$. However, the behaviour is reversed between the two data sets. While the relative mass deviation changes sign for data set II and remains negative for data set I, the relative temperature deviation displays the opposite trend: it changes sign for data set I and remains negative for data set II. Therefore, two-WHs are the most affected by the halo contribution, followed by one-WHs and, finally, RBHs.

In order to investigate how the mass $M(S)$ of the RBH responds to variations in $(q_H, V_c, a)$, we define three new parameters: $\Phi_{q_H}=\frac{\partial M}{\partial q_H}$, $\Phi_{V_c}=\frac{\partial M}{\partial V_c}$, and $\Phi_{a}=\frac{\partial M}{\partial a}$. From Eq.~(\ref{f_massa_entropia}) we obtain
\begin{align}
\Phi_{q_H}&=\frac{q_H}{2}\sqrt{\frac{\pi}{S_{q_H}}}
\left[1-\frac{2V_c^2S_{q_H}\left(3\pi a^2+S_{q_H}\right)}{\left(\pi a^2+S_{q_H}\right)^2}\right] \ ,\label{A_q} \\
\Phi_{V_c}&=-\frac{2V_cS_{q_H}^{3/2}}{\sqrt{\pi}\left(\pi a^2+S_{q_H}\right)} \ ,\label{Avc} \\
\Phi_a&=\frac{2\sqrt{\pi}aV_c^2S_{q_H}^{3/2}}{\left(\pi a^2+S_{q_H}\right)^2}\ ,\label{Aa}
\end{align}
where $S_{q_H}=\pi q_{H}^2+S$.

\begin{figure}[t!]
\includegraphics[width=\linewidth]
{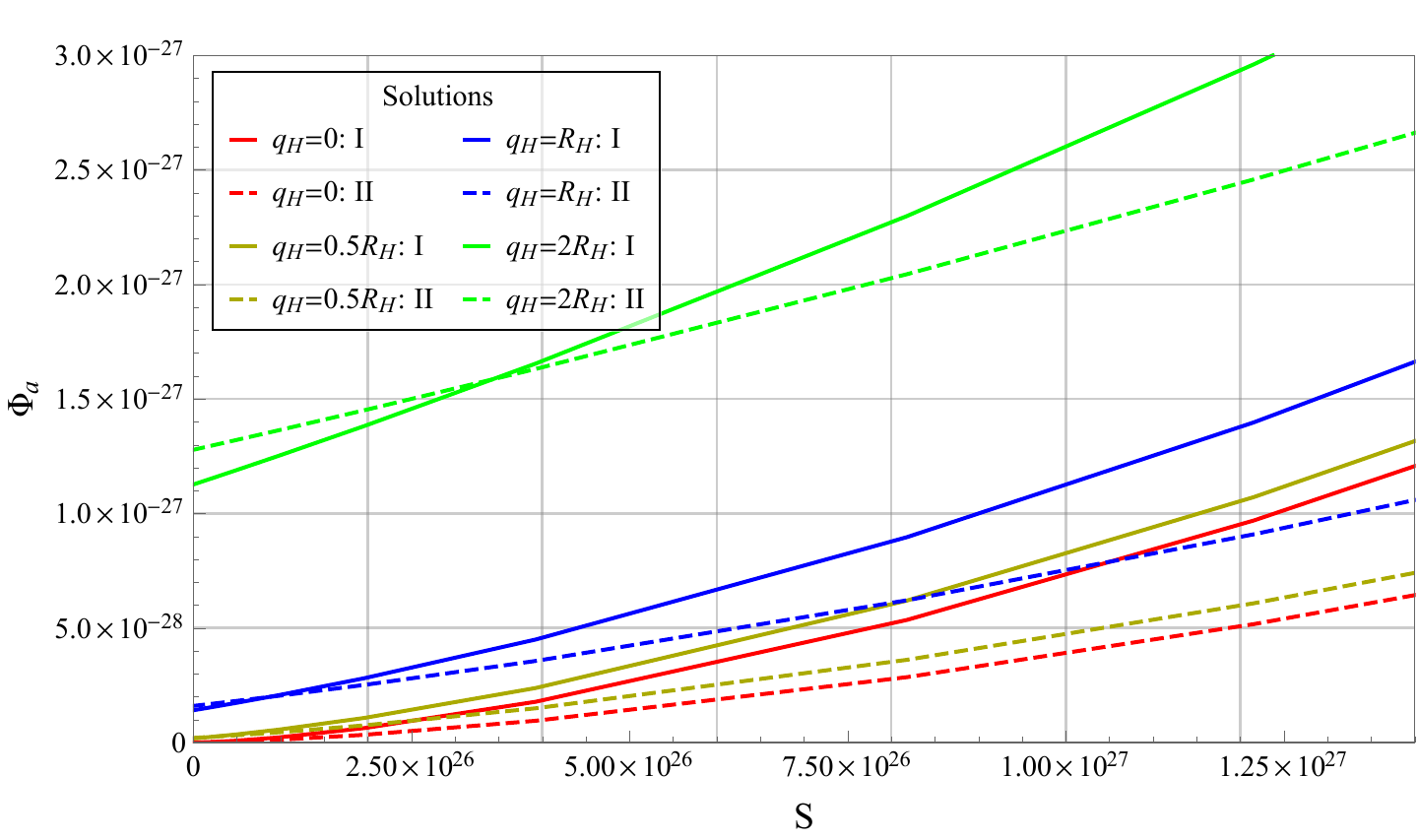}
\caption{Graphical representation of the parameter $\Phi_{q_H}$ as a function of entropy, $\Phi_{q_H}/\Phi_{q,BB} - 1$, under some values of the parameter $q_H$ for the BB solution with a halo — for values of Data I and Data II — in comparison with the BB solution without a halo. The values of $q_H$ are chosen in terms of the horizon radius of the Schwarzschild-type solution with halo, $R_{H}$.
} 
\label{fig:A_qH_entropia_BB_halo}
\end{figure}

Fig.~\ref{fig:A_qH_entropia_BB_halo} shows the behaviour of the BB mass as a function of entropy, for some values of $q_H$, through the expression $\Phi_{q_H}/\Phi_{q,BB} - 1$, from Eq.~(\ref{A_q}). This expression shows that the change in mass in the presence of the halo is mainly due to the effect of the velocity $V_c$, since if $V_c\to0$ we recover the BB without halo, as expected,
\begin{equation}
    \Phi_{q_H}|_{V_c\to0}\to \Phi_{q,BB}=\frac{q}{2}\sqrt{\frac{\pi}{S_{q}}} \ .
\end{equation}
Due to the order of magnitude of $V_c$ relative to the other parameters \cite{mod1}, the contribution of the halo becomes small at low entropy values, gaining greater significance as $V_c$ increases. The following expansion highlights the role of the radius $a$ in ensuring the halo’s influence on the mass with respect to the parameter $q_H$:
\begin{equation}
    \Phi_{q_H}/\Phi_{q,BB} - 1=-\frac{2V_cq_H^2}{a^2}\frac{(3+\frac{q_H^2}{a^2})}{(1+\frac{q_H^2}{a^2})^2}+{\mathcal O}(S) \ ,
\end{equation}
where, if the halo is sufficiently distant such that $a\gg q_H$, the behaviour returns to $\Phi_{q_H}\to \Phi_{q,BB}$, becoming independent of $q_H$.

Fig.~\ref{fig:A_qH_entropia_BB_halo} illustrates the differing behaviours of the mass according to Data I and II, under some values of the parameter $q_H$.

\begin{itemize}

\item For Data I: the solution exhibits a decrease in the mass response to entropy as the parameter $q_H$ varies, with the most pronounced effect observed for the RBH class, followed by one-WH and two-WH.

\item For Data II: the solution exhibits an increase in the mass response, with diminishing intensity as $q_H$ increases. This suggests that RBHs with a halo have their mass more strongly affected by changes in parameter $q_H$, whereas two-WHs with a halo experience a smaller influence.

\end{itemize}

\begin{figure}[t!]
\includegraphics[width=\linewidth]
{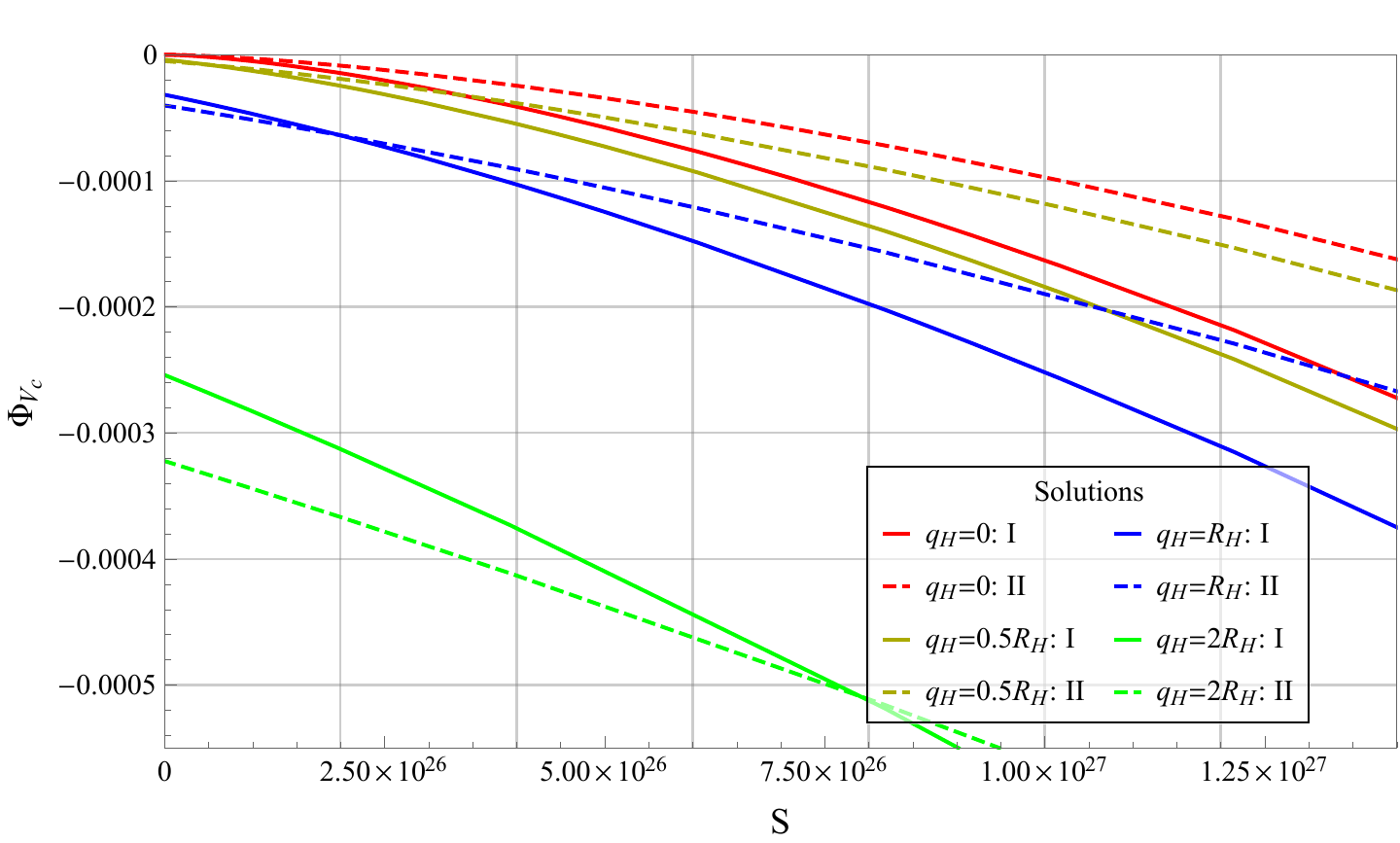}
\caption{Graphical representation of the parameter $\Phi_{V_c}$ as a function of entropy under variation of the parameter $q_H$ for the BB solution with a halo — for values of Data I and Data II. The values of $q_H$ are chosen in terms of the horizon radius of the Schwarzschild-type solution with halo, $R_{H}$.
} 
\label{fig:A_qV_entropia_BB_halo}
\end{figure}

Fig.~\ref{fig:A_qV_entropia_BB_halo} shows the behaviour of the BB mass as a function of the entropy, under variations of $V_c$, through the expression $\Phi_{V_c}$ from Eq.~(\ref{Avc}). The negative sign of this expression, $\Phi_{V_c} < 0$, indicates that, for fixed values of $(a, q_H, S)$, variations in $V_c$ and $M$ occur in opposite directions, $\delta V_c \propto -\delta M$. This suggests that, in order to preserve the properties of the system in this scenario, an increase (or decrease) in the halo velocity must be accompanied by a decrease (or increase) in the mass. That is, from Eq.~(\ref{S_definicao}), maintaining a fixed entropy corresponds to keeping the event horizon radius constant, thereby requiring a balance between $V_c$ and $M$ as a response to the halo’s influence on the spacetime curvature. Furthermore, Fig.~\ref{fig:A_qV_entropia_BB_halo} also shows that the magnitude of $\Phi_{V_c}$ increases with increasing $q_H$, highlighting the correlation between $M$ and $V_c$, which becomes more pronounced as $q_H$ increases.

From Eq.~(\ref{Avc}), under an expansion for small values of entropy,
\begin{equation}
    \Phi_{V_c}=-\frac{2V_cq_H^3}{\Sigma(a)^2}+\mathcal{O}(S) \ ,
\end{equation}
we find that the regularization factor $q_H$ becomes dominant. This structural parameter then governs the sensitivity of the mass response to the halo velocity. From Eq.~(\ref{S_definicao}), taking $S$ to be very small corresponds to considering a very small event horizon, which leads to a solution with properties close to those of a minimal or extremal BH \cite{Preskill:1991tb,Ghosh:1995rv}, no longer being controlled by the size of the event horizon.

\begin{figure}[t!]
\includegraphics[width=\linewidth]
{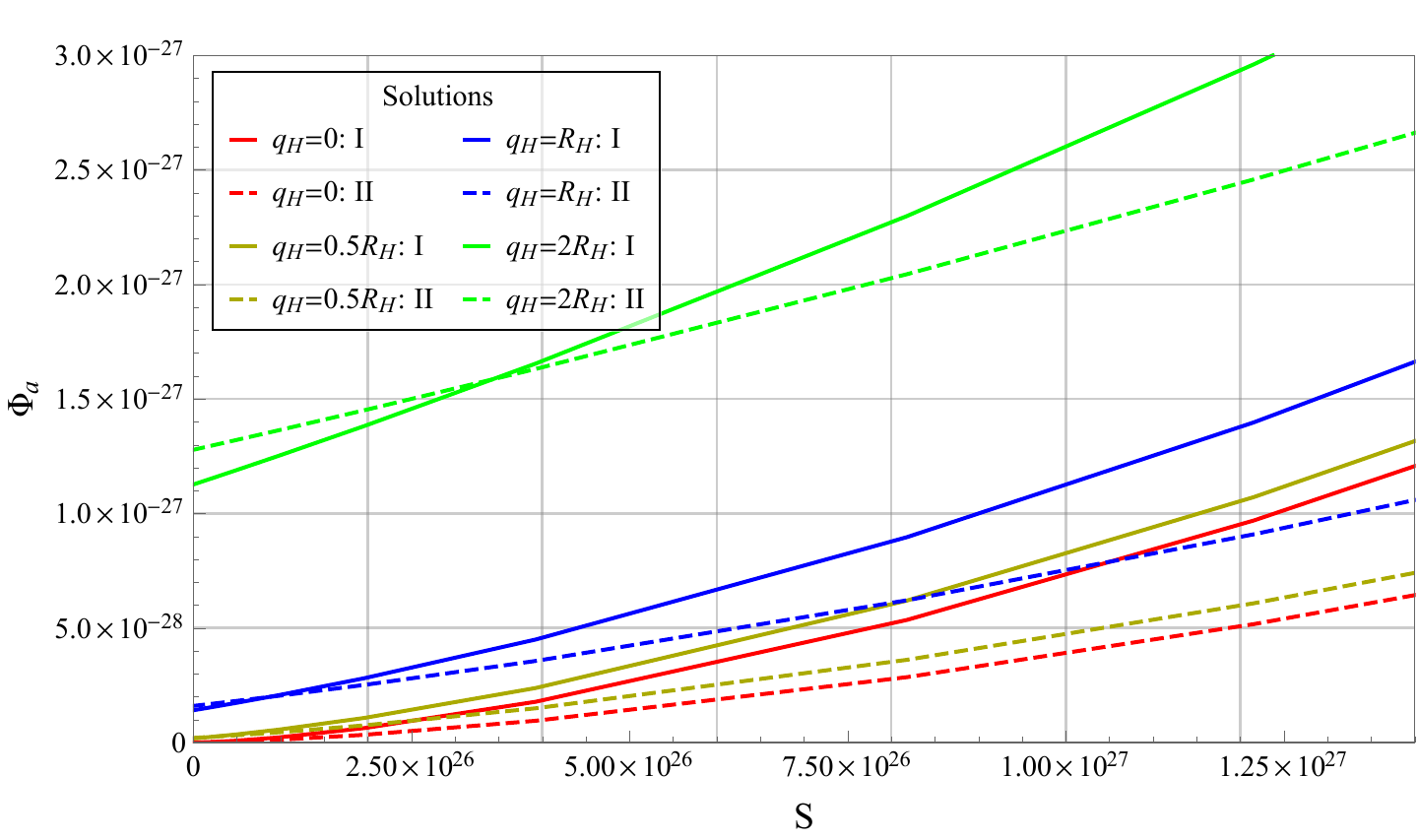}
\caption{Graphical representation of the parameter $\Phi_{a}$ as a function of entropy under variation of the parameter $q_H$ for the BB solution with a halo — for values of Data I and Data II. The values of $q_H$ are chosen in terms of the horizon radius of the Schwarzschild-type solution with halo, $R_{H}$.
} 
\label{fig:A_a_entropia_BB_halo}
\end{figure}

Fig.~\ref{fig:A_a_entropia_BB_halo} shows the behaviour of the BB mass as a function of the entropy under variations of the radius $a$, through the expression $\Phi_a$ from Eq.~(\ref{Aa}). The positive sign of this expression, $\Phi_a > 0$, indicates that, for fixed values of $(V_c, q_H, S)$, an increase in the radius $a$ — making the halo more diffuse — requires an increase in the mass $M$ to maintain the value of the event horizon radius, $\delta a \propto \delta M$. This means that the halo’s contribution to the spacetime curvature becomes more broadly distributed, making it less effective in regions close to the BH center (at $r = 0$), which in turn requires compensation through an increase in mass to preserve the same event horizon, since the entropy is held fixed. This mass response in relation to the halo radius becomes more pronounced as the parameter $q_H$ increases.

By performing an expansion of Eq.~(\ref{Aa}) around small values of $S$,
\begin{equation}
    \Phi_a=\frac{2q_H^3V_c^2a}{\Sigma(a)^2}+\mathcal{O}(S) \ ,
\end{equation}
in the relation between $M$ and $a$, an independence from the size of the event horizon emerges. The behaviour is then governed solely by the structural parameters, with greater emphasis on $q_H$. Fig.~\ref{fig:A_a_entropia_BB_halo} highlights the increasing prominence of $\Phi_a$ as $q_H$ increases.

There exists a value of entropy $S_a$ corresponding to a maximum of $\Phi_a$, given by $S_a = \pi(3a^2 - q_H^2)$. Using the constraint $S_a = \pi r_{h,a}^2 > 0$, where $r_{h,a}$ is the horizon in terms of the halo, we find that the greatest sensitivity of the mass to variations in the halo radius occurs under the condition
\begin{equation}
    \Sigma(r_{h,a})^2=3a^2 \ ,
\end{equation}
such that $a^2>q^2_H/3$. For this behaviour to be within the RBH regime, where $q_H < R$, we consider $a \gg q_H$, given that for Data I and II we have $a \sim 10^7 R$. Then, $S_a \sim \pi a^2 \to r_{h,a} \sim a$, suggesting that the halo is located close to the horizon $r_{h,a}$.

Although we present graphically the behaviour of the thermodynamic potentials in low-entropy regimes, i.e., close to zero, we emphasise that such regimes mark the limit of the semiclassical description based on Hawking radiation, where quantum gravitational effects begin to dominate the system’s dynamics. This indicates that the presence of the DM halo does not compromise the existence of a regular geometrical scenario, even where the thermodynamic description breaks down.

We also obtain the Smarr formula by rewriting the mass function, Eq.~(\ref{f_massa_entropia}), using its property as a homogeneous function, in which we perform a transformation of the form $M(S, V_c, a, q_H) \to M(y^{c_1}S, y^{c_2}V_c, y^{c_3}a, y^{c_4}q_H)$, such that
\begin{equation}
   M(y^{c_1}S,y^{c_2}V_c,y^{c_3}a,y^{c_4}q_H)=y^{1/2}M(S,V_c,a,q_H) \ ,
\end{equation}
for $c_1=1,c_2=0,$ and $c_3=c_4=c_1/2$, such that
\begin{equation}
    nM=c_1\frac{\partial M}{\partial S}S+c_2\frac{\partial M}{\partial V_c}V_c+c_3\frac{\partial M}{\partial a}a+c_4\frac{\partial M}{\partial q_H}q_H \ ,
\end{equation}
then $n = 1/2$, implying that
\begin{equation}
    M=2ST+a\phi_a+q_H \phi_{q_H} \ ,
\end{equation}
resulting in the first law of thermodynamics, as
\begin{equation}
dM=TdS+\phi_ada+\phi_{q_H}dq_H \ .
\end{equation}

\section{Charged Solution under Nonminimal Coupling}\label{sec:electrodynamic}

In \cite{bounceLED} a non-minimal framework to obtain BB-type solutions in GR was introduced via the the interaction of a scalar field $\varphi$ and a nonlinear electromagnetic field $F_{\mu\nu}$ through a non-minimal coupling function $W(\varphi)$. This scalar–electrodynamic coupling is described by
\begin{eqnarray}
\mathcal{L}_i(\varphi,F)=W(\varphi)\mathcal{L}(F) \ ,\label{acoplamento}
\end{eqnarray}
where $\mathcal{L}(F)$ is the Lagrangian density wich depends on the electromagnetic invariant $F=\frac{1}{4}F^{\mu\nu}F_{\mu\nu}$. In this formalism, there are three main ingredients: the coupling function $W(\varphi)$, the nonlinear electromagnetic Lagrangian $\mathcal{L}(F)$, and the scalar potential $V(\varphi)$. These functions determine the structure of the energy–momentum tensor and, consequently, the resulting spacetime geometry. We shall now use this formalism in order to re-interpret and analyze the structure of the BB-type solutions discussed in the sections above. 

In what follows, two main cases will be presented depending on the choice of charge. For both, we adopt $\epsilon(\varphi) = -1$, characterizing the scalar field as of the phantom type. Furthermore, we emphasize that, in the limit $q \to 0$, the contributions associated with the electromagnetic field vanish, leaving only the effective contribution of the DM halo, $T^{0}_{\ 0}\big|_{q\to0}\to\rho_{DM}$, included in the scalar potential $V(\varphi)$, as the dominant gravitational source. Since the interaction of DM  occurs exclusively through gravity, it is incorporated into the field equations via the energy–momentum tensor, without any direct coupling to the scalar or electromagnetic field, whether in the magnetically or electrically charged case. This follows from the fact that the halo’s contribution lies solely within the realm of a gravitational source. Such a separation implies that any modification induced by the presence of the halo is of a purely gravitational nature, affecting the geometry of space-time.

For completeness, the full action, the field equations, and the explicit expressions for $\mathcal{L}(r)$, $\mathcal{L}_F(r)$, and $V(r)$ are presented in Appendix~\ref{sec:appendix_A}.

\subsection{Particular solution}

Given the line element we are using, from Eq.~(\ref{metrica_geral}), and the area function $\Sigma(r) = \sqrt{r^2 + q_H^2}$, the function $\epsilon(r)$ is directly independent of the metric function $A(r)$, which simplifies the calculations. Therefore, as mentioned previously, we set $\epsilon(r) = -1$ (which characterizes a phantom field) in order to determine $\varphi(r)$, such that 
\begin{equation}
\frac{q^2_H}{\kappa^2\varphi'(r)^2\left(r^2+q_H^2\right)^2}=1 \ ,
\end{equation}
resulting in 
\begin{equation}
\varphi(r)=\frac{1}{\kappa}\tan^{-1}\left(\frac{r}{q_H} \right) \ ,\label{funcao_phi}
\end{equation}
where, asymptotically, the range of the scalar field is given by
\begin{equation}
    -\frac{\pi}{2\kappa}<\varphi<\frac{\pi}{2\kappa} \ .
\end{equation}

In this context, the term $q_H$ is interpreted as a magnetic charge ($q_H \to Q_m$) or as an electric charge ($q_H \to Q_e$). Similarly, for comparison, in the case without a SV halo we have that $q$ corresponds to the magnetic charge $q\to q_m$ or the electric charge $q\to q_e$. This mathematical formalism will be employed in the following subsections in the analysis of the Simpson–Visser solution (a BB-type solution) with a DM halo for both purely magnetic and purely electric cases.


\subsubsection{Black Bounce Solution with Dark Matter Halo and Magnetic Charge}

In this model, we consider $q_H$ exclusively as a magnetic charge $Q_m$, i.e. $q_H = Q_m$. For comparison, in the scenario without a halo, we adopt the notation of magnetic charge $q_m$, i.e. $q = q_m$. Substituting Eqs.~(\ref{funcao_phi}) and $\epsilon(r) = -1$ into the Lagrangian (\ref{L_definida_mag}), we obtain its form as a function of $W(r)$, given by
\begin{eqnarray}
\mathcal{L}(r)&=&- \frac{1}{\kappa^2}\int\frac{1}{\Sigma(r)^4W(r)^2}\Big\{W'(r)\Sigma(r)^2[(rA(r))'-1]\nonumber \\
	&& -2Q_m^2W(r)A'(r)\Big\} dr-\frac{(rA(r))'-1}{\kappa^2\Sigma(r)^2W(r)} \ .\label{Lag02_mag}
\end{eqnarray}
Aiming to employ LED, we therefore model the derivative of the Lagrangian as $\mathcal{L}_{F} = 1$ through Eq.~(\ref{LF_magnetic}), obtaining the expression for $W(r)$ as
\begin{eqnarray}
W(r)=\frac{\Sigma(r)^2}{2Q_m^2\kappa^2}\left[A''(r)\Sigma(r)^2-2A(r)+2\right] \ , \label{W(r)_mag_explic}
\end{eqnarray}
which indeed returns Eq.~(\ref{Fmag}) when we substitute Eq.~(\ref{W(r)_mag_explic}) into Eq.~(\ref{Lag02_mag}), i.e., as expected we obtain $\mathcal{L}(F) = F$, satisfying the identity
\begin{equation}
    \mathcal{L}_{F}-\frac{\partial\mathcal{L}}{\partial r}\left(\frac{\partial F}{\partial r}\right)^{-1}=0\ ,\label{eq.consistencia}
\end{equation}
which ensures the consistency of the obtained solution.

Now, we consider the metric function $A(r)$ as given in Eq.~(\ref{sol_halo_BB}), which includes the contribution of the halo. In this context, we have that $V(r)$, $\mathcal{L}(r)$, and $W(r)$ are respectively given by Eqs.~(\ref{V_definicao_mag_end}),~(\ref{Lag02_mag}),~and~(\ref{W(r)_mag_explic}). To synthesize these expressions, we write $\Sigma_m=\Sigma_m(r)=\sqrt{Q_m^2+r^2}$.  In such a case, the potential $V(r)$ takes the form
\begin{eqnarray}
    V(r)&=&-\Bigg\{V_{BB}(r)\Big|_{Q_m}+\frac{V_c^2[a^2(Q_m^2+5r^2)+\Sigma_m^4]}{24\pi(a^2+\Sigma_m^2)^3}\Bigg\}\nonumber \\
	&&\times\frac{h_3+12V_c^2\Sigma_m^3(mQ_m^2h_1-V_c^2\Sigma_m^5h_2)}{4V_c^4\Sigma_m^{10}[a^2(Q^2_m+5r^2)+\Sigma_m^4]^2-h_3}\ ,\label{V_m_final_halo}
\end{eqnarray}
where 
\begin{eqnarray*}
	h_1&=&3a^{10}(Q_m^2+3r^2)+a^8\Sigma_m^2(13Q^2_m+29r^2)\nonumber \\ && +a^6\Sigma_m^4(22Q^2_m+34r^2)+18a^4\Sigma_m^8\nonumber \\ &&+a^2\Sigma_m^8(7Q^2_m+5r^2)+\Sigma_m^{12} \ ,
		\\
	h_2&=&2a^6(Q^2_m+3r^2)(Q^2_m+5r^2)+a^4\Sigma_m^4(5Q^2_m+21r^2)\nonumber \\ && +4a^2\Sigma_m^6(Q_m^2+2r^2)+\Sigma_m^{10}   \ ,
		\\
	h_3&=&9m^2Q_m^4(a^2+\Sigma_m^2)^6    \ ,
\end{eqnarray*}
and
\begin{equation}
    V_{BB}(r)\Big|_{Q_m}=\frac{mQ_m^2}{16\pi \Sigma^5_m} \ . \label{VBB_expre}
\end{equation}
Eq.~(\ref{VBB_expre}) refers to the case without a halo, when $V_c\to0$ we have $V(r)\to V_{BB}(r)$, and likewise $Q_m\to q_m$.

The Lagrangian retains its expression as presented in Eq.~(\ref{Fmag}), except for the use of $Q_m$ instead of $q_m$, i.e. Explicitly, we have
\begin{equation}
\mathcal{L}(r)=F=\frac{Q_m^2}{2\Sigma_m^4}\ .\label{L_m_fim_halo}
\end{equation} 
The fact that $\mathcal{L}(r)$ retains the same form is due to its independence from the metric function $A(r)$. 

For $W(r)$ we have 
\begin{align}
&W(r)=W_{BB}(r)\Big|_{Q_m}+\frac{V_c^2\Sigma_m^4}{4\pi Q_m^2(a^2+\Sigma_m^2)^3}\nonumber \\
	&\qquad \times\Big[Q^2_m\left(a^2+Q_m^2\right)+r^2(5a^2+2Q_m^2
	+r^2)\Big] \ , \label{W_m_final_halo}
\end{align}
where
\begin{equation}
    W_{BB}(r)\Big|_{Q_m}=\frac{3m}{8\pi \Sigma_m} \ .\label{WBB_expre1}
\end{equation}
Eq.~(\ref{WBB_expre1}) corresponds to the limit $W(r) \to W_{BB}(r)$ when $V_c \to 0$, in which we take $Q_m \to q_m$.

\begin{figure}[t!]
\includegraphics[width=\linewidth]
{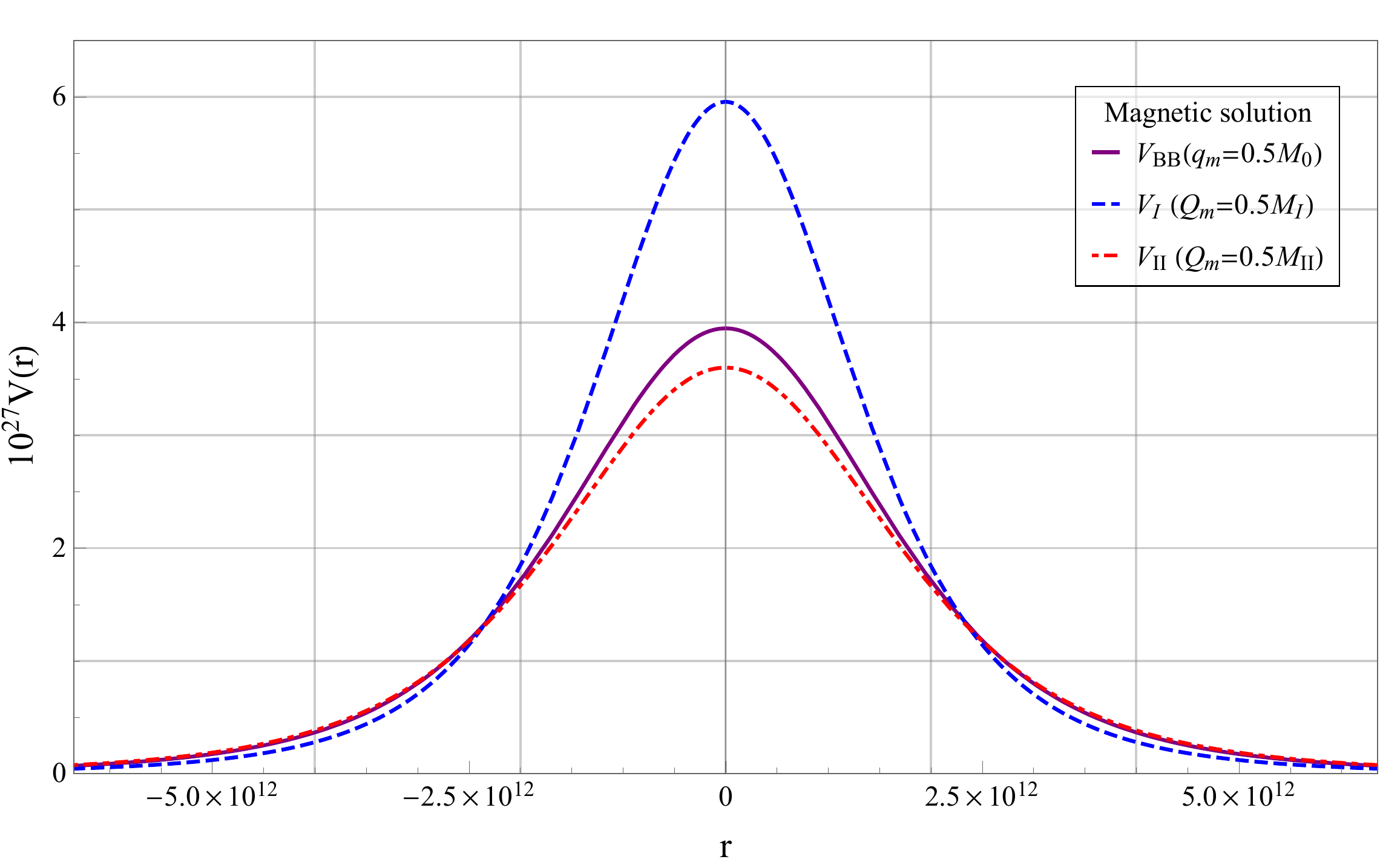}
\caption{Graphical representation of the radial behaviour of the scalar field potential $V(r)$, from Eq.~(\ref{V_m_final_halo}), using a value of the magnetic charge $0.5m$ for the BB solution with a halo — for values of Data I and Data II. For comparison, we include the case without a halo $V_{BB}(r)$ from Eq.~(\ref{VBB_expre}).
} 
\label{fig:27V_magnetic}
\end{figure}

Fig.~\ref{fig:27V_magnetic} shows that the potential $V(r)$ is modified in the presence of the halo, i.e., given the contribution of the halo for the values of Data I and II, the potential $V(r)$ exhibits different behaviour. When we consider the largest values of the halo velocity and distance (red dashed line), corresponding to $(a_{II}, V_{cII}, m_{II})$, the central region exhibits a reduction in the potential, which may suggest that part of the structural role required for the existence of the bounce is effectively carried out by the halo itself. In contrast, for smaller parameter values (blue line), corresponding to $(a_I, V_{cI}, m_I)$, a stronger contribution from the potential is required to balance the bounce, possibly due to an enhanced response associated with increased regularity.

\begin{figure}[t!]
\includegraphics[width=\linewidth]
{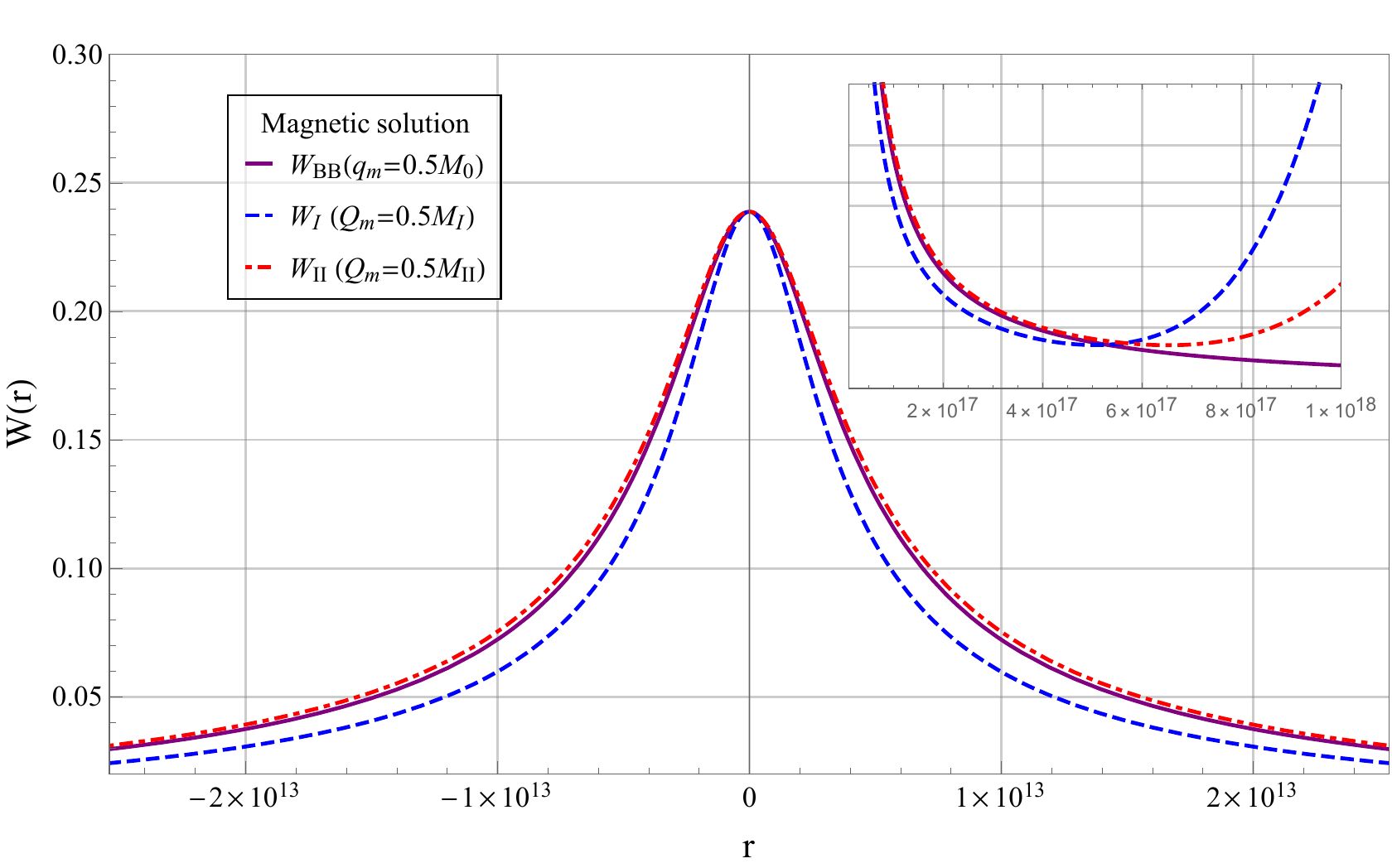}
\caption{Graphical representation of the radial behaviour of the coupling function $W(r)$, from Eq.~(\ref{W_m_final_halo}), using a value of the magnetic charge $0.5m$ for the BB solution with a halo — for values of Data I and Data II. For comparison, we include the case without a halo $W_{BB}(r)$ from Eq.~(\ref{WBB_expre1}).
} 
\label{fig:W_magnetic}
\end{figure}

Fig.~\ref{fig:W_magnetic} presents the behaviour of $W(r)$. In the central region, this coupling function is only subtly modified by the presence of the halo. From Eq.~(\ref{W_m_final_halo}), noting that $a \sim m \times 10^8$ for Data I and II, the halo contribution to $W(r = 0)$ is of the order $\sim V_c^2 \times 10^{-8}$. However, for $r \neq 0$, the halo does play a role: it acts significantly for Data I, where the decrease of the parameters $(V_c, a)$; and more subtly for Data II, where the increase of $(V_c, a)$. Furthermore, the presence of the halo results in a global minimum of $W(r)\big|_{r=\varrho}$, located between the horizon and the halo radius, $r_h < \varrho < a$, expressing the maximum suppression of the interaction between $\mathcal{L}$ and $\varphi$.

\begin{figure}[htb!]
\includegraphics[width=\linewidth]
{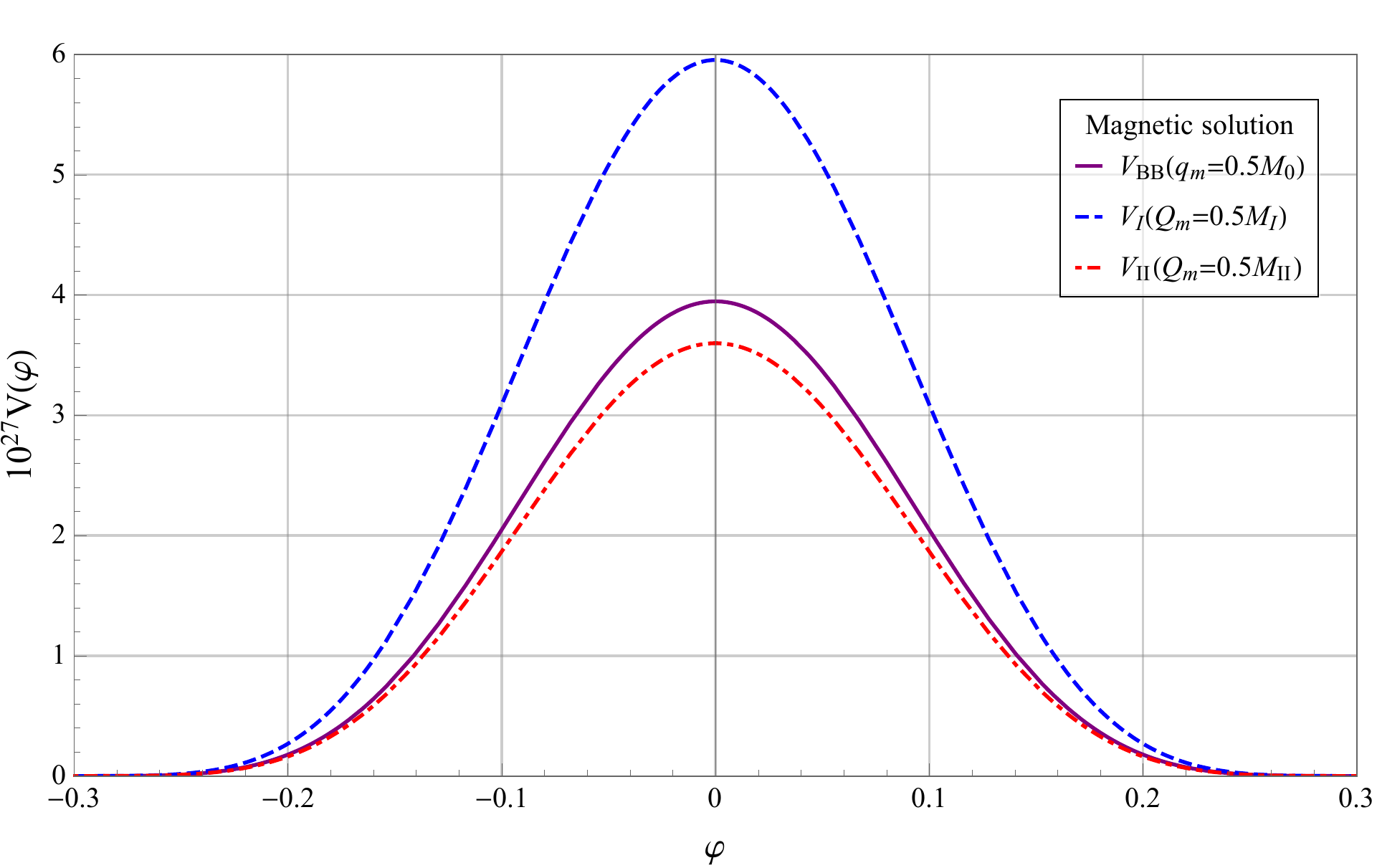}
\caption{Graphical representation of the scalar field potential $V(\varphi)$, from Eq.~(\ref{potential_V_phi}), as a function of the scalar field $\varphi$, using a value of the magnetic charge $0.5m$ in the BB solution with a halo — for values of Data I and Data II. For comparison, we include the case without a halo, $V(\varphi)\big|_{V_c=0}=V_{BB}(\varphi)$.
} 
\label{fig:27V_magnetic_phi}
\end{figure}

From Eq.~(\ref{funcao_phi}), which allows us to write $r(\varphi)$, we can define the potential, Eq.~(\ref{V_m_final_halo}), in terms of the scalar field $\varphi$, as
\begin{align}   V(\varphi)=&\Big[9m^3Q_m^6\left(a^2+Q_m^2x^2\right)^9+x_1V_c^2+x_2V_c^4
	\nonumber \\
	& +x_3V_c^6\Big]\Big[18m^2Q_m^9x^5\left(a^2+Q_m^2x^2\right)^9\nonumber \\
	&-x_4V_c^4\Big]^{-1}\label{potential_V_phi},
\end{align}
where $x=x(\varphi)=\sec(\kappa\varphi)$, and, for simplicity, we also take: 
\begin{eqnarray*}
	x_1&=&18m^2Q_m^9x^3(a^2+Q_m^2x^2)^6(-4a^4+6a^4x^2\nonumber \\ && +3a^2Q_m^2x^4+Q_m^4x^6) \ ,  
	\\
	x_2&=&-4mQ_m^{16}x^{10}(a^2+Q_m^2x^2)^3(-4a^2+5a^2x^2\nonumber \\ && +Q_m^2x^4)^2  \ ,
	\\
	x_3&=&-8Q_m^{19}x^{13}(-4a^2+5a^2x^2+Q_m^2x^4)^2(-4a^4\nonumber \\ && +6a^4x^2+3a^2Q_m^2x^4+Q_m^4x^6) \ ,    
	\\
	x_4&=&8\kappa^2Q_m^{19}x^{15}(a^2+Q_m^2x^2)^3(-4a^2+5a^2x^2\nonumber \\ && +Q_m^2x^4)^2   \ .
\end{eqnarray*}
Fig.~\ref{fig:27V_magnetic_phi} illustrates the behaviour of the potential $V(\varphi)$. The halo contribution is more significant for Data Set I, that is, when smaller values of its parameters are considered.

Similarly, for the coupling function $W(\varphi)$ we have
\begin{align}
W(\varphi)=\frac{2V_c^2Q_m^4x^4\left(5a^2x^2-4a^2+Q_m^2x^4\right)}{\kappa^2\left(a^2+Q_m^2x^2\right)^3}+\frac{3m}{\kappa^2Q_mx} \ ,\label{W_phi_mag}
\end{align}
from which we have that $W(\varphi)\big|_{V_c=0}=W_{BB}(\varphi)$. When using the parameter values from both sets, the difference between the case without a halo and the halo case is of low order, $W_{I,II}(\varphi) - W_{BB}(\varphi) \sim 10^{-16}$, due to the order of magnitude of the BH mass values. This difference becomes more pronounced as the BH mass decreases. 

In this scenario with $Q_m \neq 0 $, the Lagrangian retains its simple structure, given its dependence on angular components. In this configuration, the DM halo contributes to the component $T^{0}_{\ 0}$, which, although the scalar and magnetic fields do not interact with DM, results in an influence on the energy distribution of the system, indicating the manifestation of DM.

\subsubsection{Black Bounce Solution with Dark Matter Halo and Electric Charge}

In this model with a halo, we consider $q_H$ as an electric charge $Q_e$, i.e. $q_H = Q_e$. For future comparisons, in the scenario without a halo, we adopt the notation of electric charge $q_e$, i.e. $q = q_e$. Using the definition of the field $\varphi(r)$ from Eq.~(\ref{funcao_phi}) and $\epsilon(r) = -1$ in the Lagrangian (\ref{L_definida_elet}), we obtain its form as a function of $W(r)$, given by
\begin{eqnarray}
\mathcal{L}(r)&=&\int\frac{1}{2\kappa^2\Sigma(r)^4W(r)^2}\Big\{4Q_e^2W(r)A'(r)\nonumber \\
	&&-W'(r)\Sigma(r)^2[2rA'(r)+\Sigma(r)^2A''(r)] \Big\} dr\nonumber \\
	&&-\frac{1}{2\kappa^2W(r)}\left[A''(r)+\frac{2rA'(r)}{\Sigma(r)^2}\right] \ . \label{L_def_elec_end}
\end{eqnarray}

Similarly to the magnetic case, we make use of LED by modeling the derivative of the Lagrangian as $\mathcal{L}_{F}=1$ through Eq.~(\ref{LF_elect_}), and thus obtain the expression for $W(r)$ as
\begin{equation}
W(r)=\frac{2\kappa^2Q_e^2}{\Sigma(r)^2}\left[\Sigma(r)^2A''(r)-2A(r)+2\right]^{-1} \ .\label{W_elec_end}
\end{equation}
Furthermore, as in the magnetic case, this electric case also satisfies Eq.~(\ref{eq.consistencia}), demonstrating the consistency of the solution.

Now, we consider the halo contribution through the metric function $A(r)$ from Eq.~(\ref{sol_halo_BB}). The functions $V(r)$, $\mathcal{L}(r)$, and $W(r)$ are obtained from Eqs.~(\ref{V_definicao91}), (\ref{L_def_elec_end}), and (\ref{W_elec_end}), respectively. Similarly to the previous case, for simplicity, we take $\Sigma_e = \Sigma_e(r) = \sqrt{Q_e^2 + r^2}$.

The potential $V(r)$ is given by
\begin{align}
&V(r)=\frac{3mQ_e^2(a^2+\Sigma_e^2)^3}{3mQ_e^2(a^2+\Sigma_e^2)^3+V_c^2z_2}\Bigg[V_{BB}(r)\Big|_{Q_e}
	\nonumber \\ 
	&
	\qquad + \frac{V_c^2z_0}{12\pi\Sigma_e^2(a^2+\Sigma_e^2)^3} -\frac{V_c^4a^2\Sigma_ez_1}{12\pi mQ_e^2(a^2+\Sigma_e^2)^6}\Bigg] \  ,\label{V_e_final_halo}
\end{align}
where
\begin{eqnarray*}
	z_0&=&3a^4(Q_e^2+3r^2)+a^2(5Q_e^4+12Q^2_er^2+7r^4)+2\Sigma_e^6,
		\nonumber \\
	z_1&=&a^{10}+2a^4\Sigma_e^2(3a^4-8Q_e^4)+a^4\Sigma_e^4(15a^2+44Q_e^2)
		\nonumber \\
	&&+2a^2\Sigma_e^6(8Q_e^2-5a^2)+2\Sigma_e^8(2Q^2_e-3a^2)-2\Sigma_e^{10}, 
		\nonumber \\
	z_2&=&2a^2\Sigma_e^5(Q_e^2+5r^2)+2\Sigma_e^9 \ .
\end{eqnarray*}
such that 
\begin{equation}
V_{BB}(r)\Big|_{Q_e}=\frac{mQ_e^2}{16\pi\Sigma_e^5} \, ,
\label{VBB_e_final_semhalo}
\end{equation}
which corresponds to the case $V_c \to 0$, where $Q_e \to q_e$.

Unlike the magnetic case, the Lagrangian responds to the metric function, and is given by
\begin{eqnarray}
    \mathcal{L}(r)&=&\mathcal{L}_{BB}(r)\Big|_{Q_e}-\frac{3V_c^2mj_1}{32\pi^2\Sigma_e(a^2+\Sigma_e^2)^6} \times
    	\nonumber \\ &&
    \times \Bigg[1 -\frac{V_c^2j_0}{12mQ^2_e\Sigma_e^5j_1}\Bigg] \ ,\label{L_e_final_halo}
\end{eqnarray}
where
\begin{eqnarray*}
	j_0&=&8 a^4 \Sigma_e^{10} [7Q_e^4+5 r^2 (2 Q_e^2-r^2)]+16 a^2 \Sigma_e^{14} (Q_e^2-r^2)
		\nonumber \\
	&&+4a^6\Sigma_e^6 [5 \Sigma_e^4 (3 a^2+4
	\Sigma_e^2)+a^4 (a^2+6\Sigma_e^2)],
		\\
	j_1&=&a^8 (Q_e^2+5 r^2)+4 a^6 \Sigma_e^2 (Q_e^2+4 r^2)
		\nonumber\\
		&&
	+6 a^4 \Sigma_e^4
	(Q_e^2+3 r^2)
		+4 a^2 \Sigma_e^6 (Q_e^2+2 r^2)+\Sigma_e^{10}.
\end{eqnarray*} 
such that
\begin{equation}
    \mathcal{L}_{BB}(r)\Big|_{Q_e}=-\frac{9m^2Q_e^2}{128\pi^2\Sigma_e^6}\, ,\label{LBB_e_final_semhalo}
\end{equation}
corresponding to the case $V_c \to 0$, where $Q_e \to q_e$.

For $W(r)$, we have
\begin{eqnarray}
W(r)&=&\Bigg\{\Big[W_{BB}(r)\big|_{Q_e}\Big]^{-1}+\frac{V_c^2\Sigma_e^4}{4\pi Q_e^2(a^2+\Sigma_e^2)^3} \times
	\nonumber \\ 
	&&\times \left[a^2\big(Q_e^2 +5r^2\big)+\Sigma_e^4\right] \Bigg\}^{-1} \ ,\label{W_e_final_halo}
\end{eqnarray}
where 
\begin{equation}
W_{BB}(r)\big|_{Q_e}=\frac{8\pi \Sigma_e}{3m} \, , \label{WBB_e_final_semhalo}
\end{equation}
which corresponds to the case $V_c \to 0$, in which $W(r) \to W_{BB}(r)$, for $Q_e \to q_e$.

\begin{figure}[t!]
\includegraphics[width=\linewidth]
{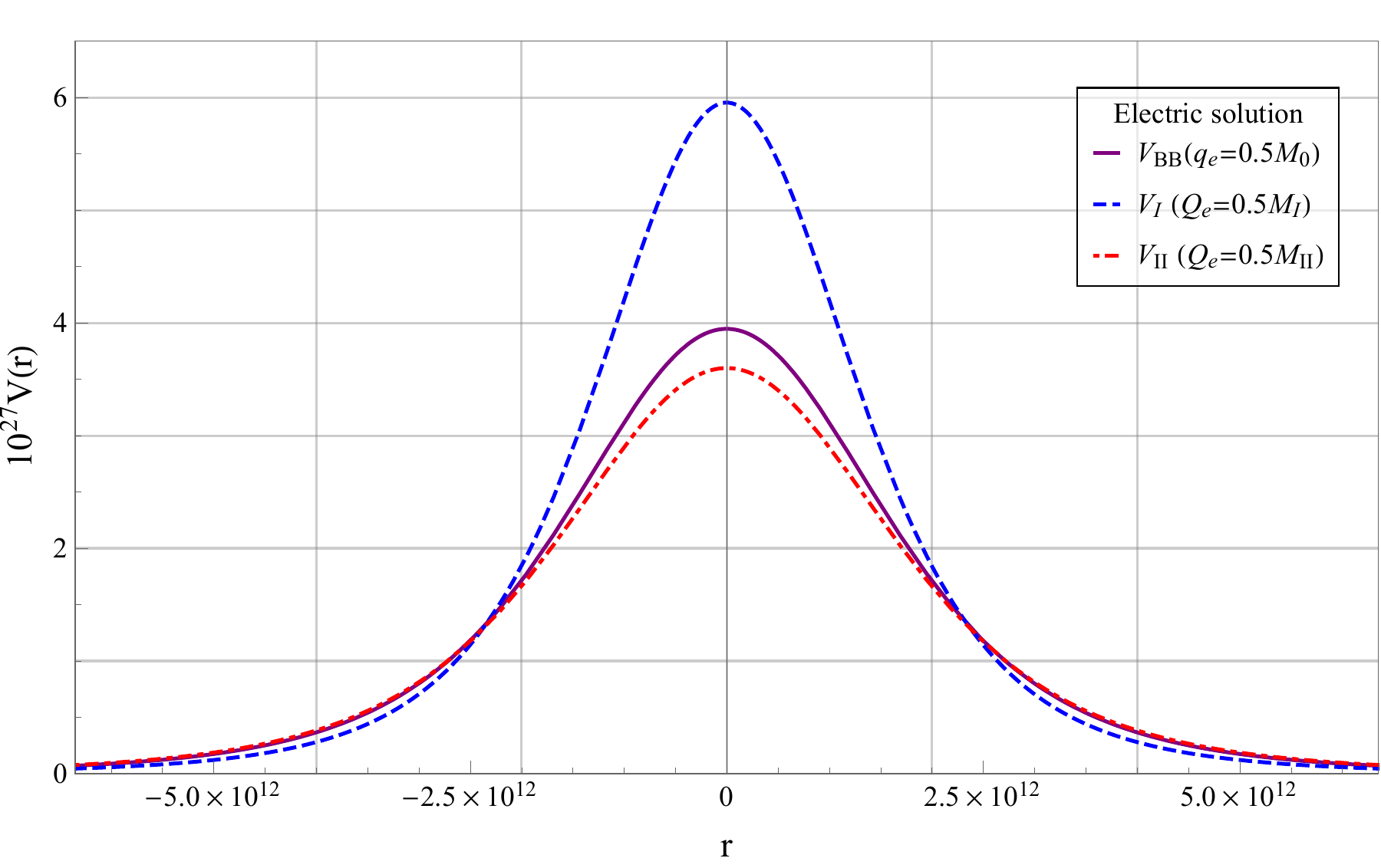}
\caption{Graphical representation of the radial behaviour of the scalar field potential $V(r)$, from Eq.~(\ref{V_e_final_halo}), using a value of the electric charge $0.5m$ for the BB solution with a halo — for values of Data I and Data II. For comparison, we include the case without a halo $V_{BB}(r)$ from Eq.~(\ref{VBB_e_final_semhalo}).
} 
\label{fig:27V_electric}
\end{figure}

Fig.~\ref{fig:27V_electric} shows that the potential $V(r)$ is modified due to the presence of the halo. The overall behaviour in this scenario is similar to the case where only magnetic charge is considered. When we choose a lower halo velocity and smaller halo radius (Data I), a significant increase in the potential occurs in the central region $r = 0$, suggesting a stronger role in sustaining the bounce. The opposite effect occurs when we choose a higher velocity and larger halo radius (Data II), indicating that part of the maintenance of regularity is taken over by another contributor, such as the halo itself.

\begin{figure}[t!]
\includegraphics[width=\linewidth]
{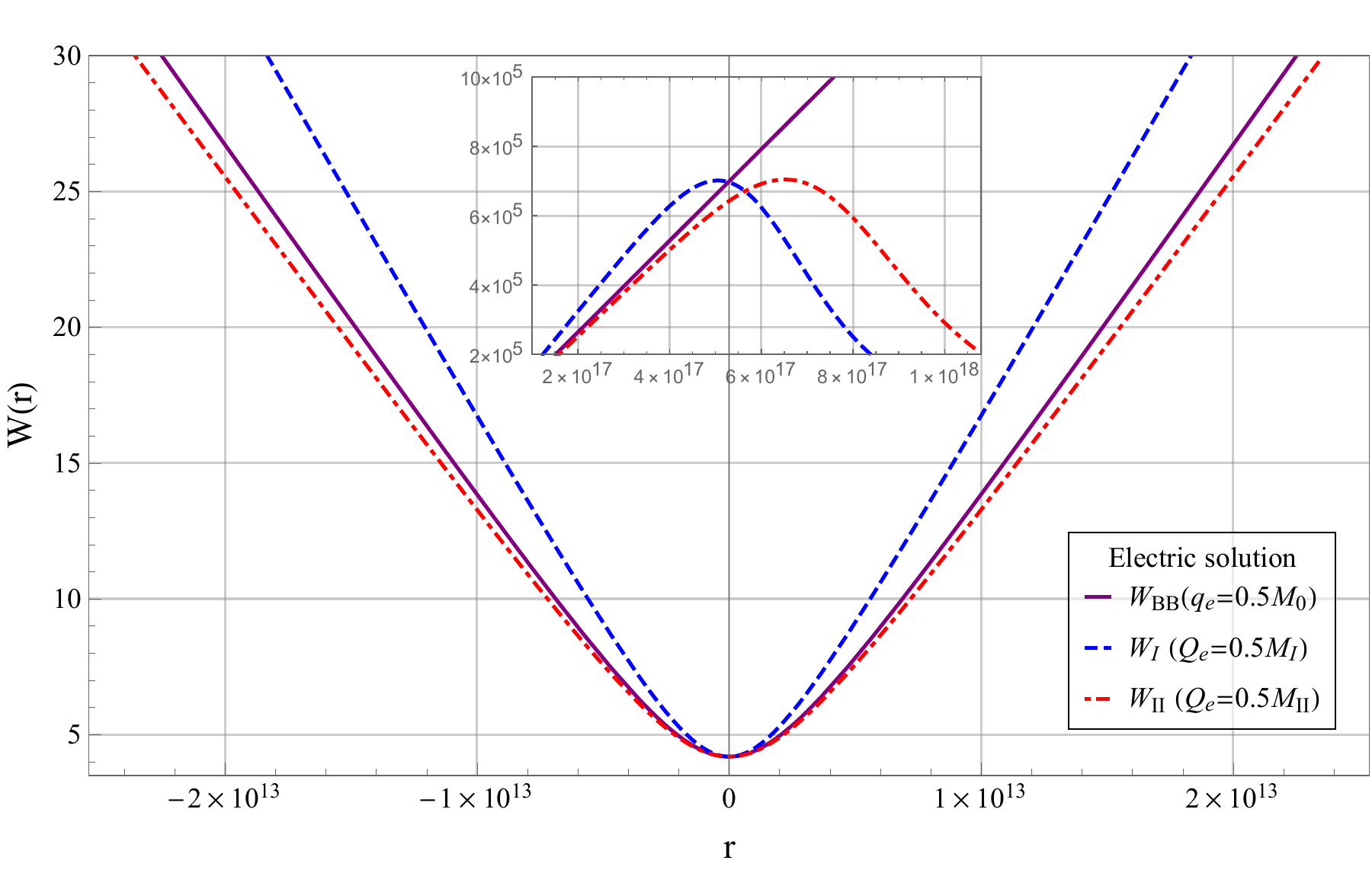}
\caption{Graphical representation of the radial behaviour of the coupling function $W(r)$, from Eq.~(\ref{W_e_final_halo}), using a 
value of the electric charge $0.5m$ for the BB solution with a halo — for values of Data I and Data II. For comparison, we include the case without a halo $W_{BB}(r)$ from Eq.~(\ref{WBB_e_final_semhalo}).
}
\label{fig:W_electric}
\end{figure}

Fig.~\ref{fig:W_electric} presents the behaviour of $W(r)$. In the central region, the halo subtly modifies the interplay between electromagnetism and the scalar field in sustaining the bounce. However, the halo contribution increases gradually as $r$ moves away from the centre. With decreasing values of the parameters $(V_c, a)$; conversely, the opposite occurs as $(V_c, a)$ increase. Despite these differing responses, in both cases there exists a global maximum of $W(r)\big|_{r=\varsigma}$, located between the horizon and the halo radius, $r_h < \varsigma < a$, expressing the maximum enhancement of the interaction between $\mathcal{L}$ and $\varphi$.

Similarly to the previous case, from Eq.~(\ref{funcao_phi}), from which we write $r(\varphi)$, we can define the potential, Eq.~(\ref{V_e_final_halo}), in terms of the scalar field $\varphi$, as
\begin{align}
    &V(\varphi)=\Bigg[\frac{3m^2(a^2+Q_e^2x^2)^3}{2Q_ex^5}+\frac{2mQ_e^2y_1}{x^2}V_c^2
    	\nonumber \\
    &\qquad \qquad -\frac{2Q_ea^2xy_2}{(a^2+Q_e^2x^2)^3}V_c^4\Bigg]\times
    \nonumber \\ 
    & \quad    \times\Bigg[\kappa^2Q_e^2\Big(3m(a^2+Q_e^2x^2)^3+2V_c^2Q_e^4x^5y_3\Big)\Bigg]^{-1},
\end{align}
where
\begin{eqnarray*}
	y_1&=&(-6a^4+9a^4x^2-2a^2Q_e^2x^2+7a^2Q_e^2x^4+2Q_e^4x^6), 
		\nonumber\\
	y_2&=&(a^{10}+6a^8Q_e^2x^2-16a^4Q_e^6x^2+15a^6Q_e^4x^4
		\nonumber \\
	&&+
	44a^4Q_e^6x^4-10a^4Q_e^6x^6+16a^2Q_e^8x^6
		\nonumber \\
	&&-6a^2Q_e^8x^8+4Q_e^{10}x^8-2Q_e^{10}x^{10}), 
		\nonumber \\
	y_3&=&Q_e( x^4 + 5 a^2 x^2-4 a^2) \, .
\end{eqnarray*}

Similarly, we obtain the coupling function in terms of the scalar field, $W(\varphi)$, as
\begin{eqnarray}
    W(\varphi)&=&\big[\kappa^2 Q_e^3 x (a^2 + Q_e^2 x^2)^3\big]\big[3mQ_e^2(a^2 + Q_e^2 x^2)^3 \nonumber \\ &&+ 
 2V_c^2 Q_e^5 x^5 (-4 a^2 + 5 a^2 x^2 + Q_e^2 x^4)\big]^{-1}\, ,
\end{eqnarray}
where, as in the magnetic case, it exhibits a low-order difference relative to the case without a halo, $W(\varphi)\big|_{V_c=0}=W_{BB}(\varphi)$, with $W_{BB}(\varphi)-W_{I,II}(\varphi)\sim 10^{-15}$.

In this scenario with $Q_e \neq 0$, unlike the magnetic case, the structure of the Lagrangian is modified due to the form of $F$, directly affecting the coupling term. Owing to this distinct reconstruction of the Lagrangian, the presence of the halo alters the geometry of the space-time, which in turn indirectly influences the behaviour of the electric field, as reflected in the form of the Lagrangian.

\section{Conclusion and discussion}\label{sec:concl}

In this work, we have proposed an implementation of the Simpson--Visser Black Bounce (BB) solution describing a black hole (BH) immersed in a dark matter (DM) halo. This model adopts the DM density profile $\rho_{DM}$ introduced in \cite{mod1}, characterized by the halo velocity $V_c$ and the critical radius $a$, which is consistent with data from the Hubble Space Telescope (HST) and globular cluster dynamics of the elliptical galaxy NGC 4649 (M60). The metric function obtained, as well as all expressions presented here within the BB framework, reduce to the standard BB solution without a halo when $V_c\to0$ (or $a\to\infty$), and to the Schwarzschild-type solution with a DM halo \cite{Lobo2025H1} when the bounce parameter vanishes, $q_H\to0$. The parameter values are taken from \cite{mod1}, which we divide into two representative datasets — Data I and Data II — in order to examine the influence of the halo on certain properties of the BB solution \cite{Simpson:2018tsi}.

The first noticeable effect of the halo appears in the behaviour of the regularization parameter $q_H$, which in the usual, non-halo case is denoted by $q$. From the horizon calculation, its domain is reduced for Data I and enlarged for Data II, indicating a direct influence of the halo-induced modifications in the mass values. Moreover, for a fixed value of $q_H$, which we consider to be of the order of the Schwarzschild radius without a halo, the metric function behaves differently depending on the dataset, suggesting that the halo's influence extends to the nature of the geometries. Despite the modifications to the domain of the regulating parameter, the classes of BB solutions retain their thresholds following the horizon radii of the Schwarzschild-type solution: in the BB case with a halo, the domain of $q_H$ follows the threshold governed by the event horizon radius arising from the Schwarzschild solution with a halo. The modification of the event horizon constitutes an implicit response to the change in the critical value of the regulator parameter in the presence of the DM halo, and can be interpreted as an essentially geometrical effect. Although it is not a direct observable, the shift in the critical value of $q_H$ may serve as a basis for discussing other observables, such as those associated with post-merger BH systems, in which horizon oscillations are related to quasi-normal modes.

The Kretschmann scalar makes explicit the role of the halo in the space-time curvature. Its general asymptotic behaviour is preserved, vanishing as one moves away from the center and converging to a finite constant at $r = 0$. This regular central value is significantly affected by the halo contribution, leading to a reduction of the curvature for Data I and an increase for Data II in the regime $q_H \lesssim R = 2M_0$; an inversion occurs when $q_H > R$. The two-WH case exhibits a predominantly amplified $K$, in contrast to the one-WH and RBH cases under Data I, with the opposite overall behaviour for Data II.

Due to the radial dependence of the halo contribution in the metric, we computed the domain of $q_H$ for Data I and II using the shadow radius $r_{sh}$. For both datasets, the shadow radius decreases with increasing $|q_H|$, and only values of $q_H$ corresponding to RBHs lie within the valid interval according to observational limits ($1\sigma$ and $2\sigma$), suggesting a limitation of the halo model in the context of imaging analysis. We further analyzed the photon trajectories by considering circular orbits through the effective potential $V_{eff}(r)$ and the radial acceleration $\alpha(r)$. While the halo does not alter the stability type of the orbit, it does affect the intensity of the potential as well as the particle's radial acceleration. A reduction in both the potential and acceleration is observed for Data I, whereas the opposite effect occurs for Data II. Numerically, we observe an increase in the angular velocity for Data I and a decrease for Data II. From the series expansion, the contributions from the halo velocity and radius become explicit, indicating that the escape dynamics depend on the extent of the halo.

Our thermodynamic analysis revealed that the mass–entropy and temperature–entropy relations respond differently for Data I and Data II. Data II presents a condition ($S = S_{0,II}$) for which $M \to M_{BB}$, effectively nullifying any modification of the mass due to the halo; Data I exhibits an analogous condition ($S = S_{0,I}$) under which $T \to T_{BB}$. We also defined thermodynamic potentials in the BB solution by considering the characteristic halo parameters. The potential $\phi_{V_c}$ indicates an opposite behaviour between variations of mass and halo velocity ($\Phi_{V_c} < 0$), while the potential $\phi_a$ shows that the variation of mass increases as the halo radius becomes larger. All thermodynamic parameters exhibit more pronounced responses as $q_H$ increases, particularly due to the low values of the velocity $V_c$. The correspondence between the halo and without-a-halo scenarios is therefore achieved either in the limit of a highly diffuse halo or when the halo velocity vanishes.

Furthermore, we extended our analysis to charged BB-type solutions under non-minimal coupling. In the purely magnetic ($Q_m \neq 0, Q_e = 0$) and purely electric ($Q_m = 0, Q_e \neq 0$) cases, the halo alters both the intensity and the radial distribution of the potential $V(r)$, the Lagrangian $\mathcal{L}(r)$, and the coupling function $W(r)$. The general expression for the scalar field $\varphi(r)$, arising from the choice of a phantom field, is preserved; however, its behaviour is subtly modified because the charged source depends on the BH mass, which responds to the presence of the halo. In a scenario with a more compact and slower halo (Data I), there is an enhancement of the Lagrangian contribution and of the influence of the potential in the central region. Conversely, for a more extended and faster halo (Data II), these effects are softened, implying that part of the geometrical regularization is effectively carried by the halo itself. The halo also affects $W(r)$ as one moves away from the center, leading to the appearance of a global maximum of the coupling located between the horizon and the halo radius.

In summary, the presence of a DM halo induces significant modifications in the geometrical structure, the dynamics of massless particles, and the thermodynamic behaviour of BB-type solutions. In the context of charged solutions, modifications arise in the roles of the phantom scalar field and the electromagnetic field, as well as in their interaction. The halo thus emerges as a structural element in sustaining regularity, rather than merely an external correction. These findings open new avenues for understanding the astrophysical relevance of DM in regimes of strong gravitational fields, suggesting that observable properties of BHs may serve as indirect probes of the DM distribution in galactic cores.

Our results are naturally framed within the different approaches considered in the literature regarding potential measurements of both direct and indirect observables that could indicate the existence of a DM halo surrounding BHs. Among these are modifications in the shadow radius, shifts in photon orbits, subtle effects in strong gravitational lensing, and possible alterations in the accretion disk dynamics. These effects arise as small but cumulative corrections to the observables predicted by GR, while still preserving the fundamental laws of gravity and thermodynamics \cite{Benkrane2025}. Another way to probe such effects is through systems in which massive objects orbit these BHs. In this context, the elliptical galaxy NGC 4649---whose density profile we have adopted here---constitutes a promising natural laboratory, since it hosts the ultracompact galaxy UCD1 orbiting a supermassive BH \cite{UCD1}. Although the halo-induced modifications are small and systematic, future observations with improved instrumental precision may enable the detection of orbital deviations, providing independent yet complementary tests for refining realistic models of supermassive BHs embedded in DM halos, such as the one studied in this work.

Future work will deepen the physical interpretation of this model through a variety of theoretical and observational approaches. These include analyzing the stability of the proposed solution under both linear and non-linear perturbations, investigating gravitational lensing in this space-time to constrain the parameters $V_c$ and $a$ through observations around supermassive BHs, and studying the Hawking radiation spectrum in this halo-influenced metric. Collectively, these research directions will help to refine the theoretical model and bridge the gap between the astrophysical observational regime and cosmology.

\acknowledgments{
FSNL acknowledges support from the Funda\c{c}\~{a}o para a Ci\^{e}ncia
e a Tecnologia (FCT) Scientific Employment Stimulus contract with reference CEECINST/00032/2018, and funding through the research grants UID/04434/2025. MER thanks Conselho Nacional de Desenvolvimento
Cient\'{\i}fico e Tecnol\'ogico (CNPq), Brazil, for partial financial support. DRG acknowledge finantial support by the Spanish National Grants PID2022-138607NBI00 and CNS2024-154444 grants, funded by MICIU/AEI/10.13039/501100011033. This study was financed in part by the
Coordena\c{c}\~{a}o de Aperfei\c{c}oamento de Pessoal de N\'{\i}vel Superior - Brasil (CAPES) - Finance Code 001.}


\appendix
\section{CHARGED SOLUTION UNDER NONMINIMAL COUPLING}
\label{sec:appendix_A}

In this appendix, we present the technical details of the model considered in Section \ref{sec:electrodynamic}, complementing the discussion provided in the body of this paper.

The electromagnetic tensor $F_{\mu\nu}$ is defined in terms of the vector potential $A_{\mu}$ as $F_{\mu\nu}=\partial_{\mu}A_{\nu}-\partial_{\nu}A_{\mu}$.  The action $\mathcal{S}$ employed comprises the Einstein-Hilbert gravitational term, a scalar field, and the non-minimal interaction term $\mathcal{L}_i(\varphi,F)$ from Eq.~(\ref{acoplamento}). This action is given by
\begin{equation}
\mathcal{S}=\int\sqrt{-g}d^4x\left[R-2\kappa^2\left(\mathcal{L}_{\varphi}(\varphi)-\mathcal{L}_i(\varphi,F)\right) \right] \ ,\label{acaoS}
\end{equation}
where $\kappa^2=8\pi$, and $\mathcal{L}_{\varphi}(\varphi)$ is the Lagrangian associated with the scalar field, which reads as
\begin{equation}
\mathcal{L}_{\varphi}(\varphi)=\epsilon(\varphi)\partial^{\mu}\varphi\partial_{\mu}\varphi-V(\varphi) \ ,
\end{equation}
where $V(\varphi)$ represents the scalar field potential. Furthermore, in this work we choose $\epsilon(\varphi) < 0$, which defines a phantom-type scalar field.

We emphasize that the contribution of DM to the action $\mathcal{S}$, Eq.~(\ref{acaoS}), is incorporated into the scalar potential $V(\varphi)$. In this way, DM does not interact directly with either the scalar field $\varphi$ or the electromagnetic field $F$, but instead acts as a gravitational source. This behaviour becomes clearer when considering the limit $q\to0$, in which the contributions associated with the electromagnetic source vanish, leaving only the DM contribution in $V(\varphi)$. Consequently, as we shall see below, this contribution manifests in the energy–momentum tensor $T^{\mu}_{ \ \nu}$, particularly in the density component, such that $T^{0}_{ \ 0}\big|_{q\to0}\to\rho_{DM}$, thereby evidencing the role of the halo.

Varying the action (\ref{acaoS}) with respect to the metric tensor $g_{\mu\nu}$, we have
\begin{equation}
G^{\mu}_{\ \nu}\equiv R^{\mu}_{\ \nu}-\frac{1}{2}\delta^{\mu}_{\ \nu}R=\kappa^2T^{\mu}_{\ \nu} \ ,\label{tensorG}
\end{equation}
such that 
\begin{equation}
T^{\mu}_{\ \nu}=W(\varphi)\mathcal{T}^{\mu}_{\ \ \nu}+\Theta^{\mu}_{\ \nu} \ ,
\end{equation}
where the energy-momentum tensor is 
\begin{equation}
\mathcal{T}^{\mu}_{\ \ \nu}=\delta^{\mu}_{\ \nu}\mathcal{L}(F)-\mathcal{L}_FF^{\mu\alpha}F_{\nu\alpha} \ ,
\end{equation}
and the energy-momentum tensor for the scalar field is as
\begin{equation}
\Theta^{\mu}_{\ \nu}=\epsilon\left(2\partial^{\mu}\varphi\partial_{\nu}\varphi-\delta^{\mu}_{\ \nu}\partial^{\sigma}\varphi\partial_{\sigma}\varphi\right)+\delta^{\mu}_{\ \nu}V(\varphi) \ .
\end{equation}

Now, by varying the action (\ref{acaoS}) with respect to the potential $A_{\mu}$ and the field $\varphi$, we obtain the following equations of motion:
\begin{equation}
\nabla_{\mu}\left[W(\varphi)\mathcal{L}_{F}(F)F^{\mu\nu}\right]=0 \ ,\label{eq1_W}
\end{equation}
and
\begin{equation}
    2\epsilon(\varphi)\square\varphi
+\nabla_{\mu}\varphi\nabla^{\mu}\varphi\frac{d\epsilon(\varphi)}{d\varphi}+\mathcal{L}(F)\frac{dW(\varphi)}{d\varphi}=-\frac{dV(\varphi)}{d\varphi} \ ,\label{eq2_W}
\end{equation}
where $\square=\nabla_{\mu}\nabla^{\mu}$, and $\mathcal{L}_{F}=\partial\mathcal{L}(F)/\partial F$.

Now, we make use of the line element from Eq.~(\ref{metrica_geral_BB}), which assumes a spherically symmetric and static solution, to obtain the non-trivial components of the Einstein tensor from Eq.~(\ref{tensorG}), which are given by:
\begin{equation}
G^{0}_{\ 0}=\frac{A(r)\Sigma'(r)^2-1}{\Sigma(r)^2}+\frac{A'(r)\Sigma'(r)+2A(r)\Sigma''(r)}{\Sigma(r)} \ ,
\end{equation}
\begin{equation}
G^{1}_{\ 1}=\frac{A(r)\Sigma'(r)^2-1}{\Sigma(r)^2}+\frac{A'(r)\Sigma'(r)}{\Sigma(r)} \ ,
\end{equation}
and
\begin{equation}
G^{2}_{\ 2}=G^{3}_{\ 3}=\frac{A'(r)\Sigma'(r)+A(r)\Sigma''(r)}{\Sigma(r)}+\frac{A''(r)}{2}\, .
\end{equation}

Although the Maxwell–Faraday tensor $F_{\mu\nu}$ has the non-vanishing components $F_{01} = -F_{10}$ and $F_{23} = -F_{32}$, corresponding to the contributions of the electric charge $q_e$ and the magnetic charge $q_m$, in this work we consider only two cases: purely electric and purely magnetic.

\subsection{Magnetic case}

In the purely magnetic case, where we set $q_e = 0$, the non-vanishing component is $F^{23} = -F^{32}$, given by
\begin{equation}
F^{23}=\frac{q_m\csc{\theta}}{\Sigma(r)^4} \ .\label{F23}
\end{equation}
Thus, the scalar $F$ is given by
\begin{equation}
    F=\frac{q^2_m}{2\Sigma(r)^4} \ .\label{Fmag}
\end{equation}
We then use Eqs.~(\ref{metrica_geral_BB}), (\ref{F23}),~and~(\ref{Fmag}) in the field equations~(\ref{tensorG})~to obtain the following equations of motion, $G^{\mu}_{\ \nu}=\kappa^2T^{\mu}_{\ \nu}$, given by:
\begin{equation}
    \kappa^2T^{0}_{\ 0}=\kappa^2\left[V(r)+\mathcal{L}(r)W(r)+A(r)\epsilon(r)\varphi'(r)^2\right]\ ,\label{T00_mag}
\end{equation}
\begin{equation}
\kappa^2T^{1}_{\ 1}= \kappa^2\left[V(r)+\mathcal{L}(r)W(r)-A(r)\epsilon(r)\varphi'(r)^2\right] \ ,\label{T11_mag}
\end{equation}
and
\begin{equation}
\kappa^2T^{2}_{\ 2}=\kappa^2T^{3}_{\ 3}=\kappa^2\left[T^{0}_{\ 0}-W(r)\frac{q_m^2\mathcal{L}_F(r)}{\Sigma(r)^4} \right] \ ,\label{T22_mag}
\end{equation}

The general expressions for $\mathcal{L}(r)$ and $\mathcal{L}_F(r)$ can be determined from Eqs.~(\ref{T00_mag}) and (\ref{T22_mag}), given by
\begin{eqnarray}
    \mathcal{L}(r)&=&-\frac{1}{\kappa^2W(r)\Sigma(r)^2}\Big\{\Sigma(r)\left[A'(r)\Sigma'(r)+2A(r)\Sigma''(r)\right]
    	\nonumber \\	&&+A(r)\Sigma'(r)^2+\kappa^2\Sigma(r)^2[A(r)\epsilon(r)\varphi'(r)^2	\nonumber \\
	&&+V(r)]-1\Big\}\ ,\label{L_definida_mag}
\end{eqnarray}
and 
\begin{eqnarray}
    \mathcal{L}_F(r)&=&\frac{\Sigma(r)^2}{2q_m^2\kappa^2W(r)}\Big\{A''(r)\Sigma(r)^2-2A(r)\big[\Sigma'(r)^2
    \nonumber \\
	&&+\Sigma(r)\Sigma''(r)\big]+2 \Big\} \ .\label{LF_magnetic}
\end{eqnarray}
Thus, with $\mathcal{L}(r)$ from Eq.~(\ref{L_definida_mag}), the definition of $\epsilon(r)$ can be obtained from Eq.~(\ref{T11_mag}):
\begin{equation}
-\frac{2A(r)}{\Sigma(r)}\left[\kappa^2\epsilon(r)\Sigma(r)\varphi'(r)^2+\Sigma''(r)\right]=0 \ .\label{eq_epsolon}
\end{equation}
From Eq.~(\ref{eq_epsolon}), we obtain $\epsilon(r)$ as
\begin{equation}
    \epsilon(r)=-\frac{\Sigma''(r)}{\kappa^2\Sigma(r)\varphi'(r)^2} \ .\label{def_epsolon}
\end{equation}

Now, in order to define the potential $V(r)$, we use Eq.~(\ref{def_epsolon}) in Eq.~(\ref{eq2_W}), and then perform the integration to obtain the following expression:
\begin{eqnarray}
V(r)&=&-W(r)\int \frac{1}{\kappa^2W(r)^2\Sigma(r)^2}\Big\{W(r)\Big[3A(r)\Sigma'(r)    \nonumber \\
	&&
    \times\Sigma''(r)+\Sigma(r)\left[2A'(r)\Sigma''(r)+A(r)\Sigma'''(r)\right]\Big]
    \nonumber \\
	&&
-W'(r)\Big[\Sigma(r)\left[A'(r)\Sigma'(r)+A(r)\Sigma''(r)\right]\nonumber \\
	&&+A(r)\Sigma'(r)^2-1\Big]\Big\}dr\, .\label{V_definicao_mag_end}
\end{eqnarray}

\subsection{Electric case}

In the purely electric case, where we set $q_m = 0$, the non-vanishing component is $F^{10} = -F^{01}$, given by
\begin{equation}
F^{10}=\frac{q_e}{W(r)\mathcal{L}_F(r)\Sigma(r)^2} \ .\label{F10}
\end{equation}
Thus, the scalar $F$ takes the form
\begin{equation}
F=-\frac{q^2_e}{2W(r)^2\mathcal{L}_F(r)^2\Sigma(r)^4} \ .\label{Felet}
\end{equation}

Then, similarly to the purely magnetic case, we substitute Eqs.~(\ref{metrica_geral_BB}), (\ref{F10}), and (\ref{Felet}) into the field equations (\ref{tensorG}), thereby obtaining the following equations of motion, $G^{\mu}_{\ \nu}=\kappa^2T^{\mu}_{\ \nu}$, given by:
\begin{eqnarray}
\kappa^2T^{0}_{\ 0}&=&\kappa^2\Bigg[\frac{q_e^2}{\mathcal{L}_F(r)W(r)\Sigma(r)^4}+A(r)\epsilon(r)\varphi'(r)^2\nonumber \\
	&&+\mathcal{L}(r)W(r)+V(r) \Bigg] \ ,\label{T00elet}
\end{eqnarray}
and,
\begin{equation}
\kappa^2T^{1}_{\ 1}=\kappa^2\left[T^{0}_{\ 0}-2A(r)\epsilon(r)\varphi'(r)^2\right]\ ,\label{T11elet}
\end{equation}
and,
\begin{eqnarray}
\kappa^2T^{2}_{\ 2}&=&\kappa^2T^{3}_{\ 3}=\kappa^2\Big[\mathcal{L}(r)W(r)+A(r)\epsilon(r)\varphi'(r)^2\nonumber \\	&&+V(r)\Big]\label{T22elet} \ .
\end{eqnarray}

The expressions for $\mathcal{L}(r)$ and $\mathcal{L}_F(r)$ can be determined from Eqs.~(\ref{T00elet}) and (\ref{T22elet}), as
\begin{align}
\mathcal{L}(r)=&-\frac{1}{2\kappa^2W(r)\Sigma(r)}\Big\{2\kappa^2\Sigma(r)[A(r)\epsilon(r)\varphi'(r)^2\nonumber \\
	&+V(r)]+2[A(r)\Sigma'(r)]'+\Sigma(r)A''(r) \Big\} \ ,  \label{L_definida_elet}
\end{align}
and,
\begin{eqnarray}
\mathcal{L}_F(r)&=&-\frac{2\kappa^2q_e^2}{W(r)\Sigma(r)^2}\Big\{\Sigma(r)^2A''(r)-2A(r)\nonumber \\	&&\times\left[\Sigma'(r)^2+\Sigma(r)\Sigma''(r)\right]+2\Big\}^{-1}\ .\label{LF_elect_}
\end{eqnarray}

Thus, having $\mathcal{L}(r)$ from Eq.~(\ref{L_definida_elet}), the definition of $\epsilon(r)$ can be obtained from the component $T^{1}_{\ 1}$ of Eq.~(\ref{T11elet}), resulting in the same expression for $\epsilon(r)$ as in the purely magnetic case, given in Eq.~(\ref{eq_epsolon}). However, the expression for the potential takes a different form, being obtained by using Eq.~(\ref{def_epsolon}) in Eq.~(\ref{eq2_W}), as
\begin{eqnarray}
V(r)&=&W(r)\int \frac{1}{2\kappa^2W(r)^2\Sigma(r)^2}\Big\{\Sigma(r)^2W'(r)A''(r)
    \nonumber \\
	&&
    -6A(r)W(r)\Sigma'(r)\Sigma''(r)-2\Sigma(r)\Big[A'(r)\times
        \nonumber \\
	&&[2W(r)\Sigma''(r)
    -W'(r)\Sigma'(r)]
        \nonumber \\
	&&-A(r)W(r)\Sigma'''(r)\Big] \Big\}dr\label{V_definicao91} \ .
\end{eqnarray}




\end{document}